\documentclass[a4paper,fleqn]{cas-sc}

\usepackage{setspace}
\usepackage{bm}
\usepackage{multirow}
\usepackage[numbers,sort&compress]{natbib}
\usepackage{graphicx}
\usepackage{float}
\usepackage{geometry}
\usepackage{enumitem}
\usepackage{amsmath}
\usepackage{times}
\usepackage{caption}
\usepackage[ruled]{algorithm2e}
\usepackage[title]{appendix}%

\usepackage{lineno}
\def\tsc#1{\csdef{#1}{\textsc{\lowercase{#1}}\xspace}}
\tsc{WGM}
\tsc{QE}
\tsc{EP}
\tsc{PMS}
\tsc{BEC}
\tsc{DE}
\begin{document}

\let\WriteBookmarks\relax
\def\floatpagepagefraction{1}
\def\textpagefraction{.001}
\shorttitle{Families of LEO-DPO Transfers}
\shortauthors{Shuyue Fu et~al.}

\title [mode = title]{Families of Transfers from circular low Earth orbit to Distant Prograde Orbit around the Moon}                      

\author[1,2]{Shuyue Fu}[orcid=0009-0001-5111-9779,style=chinese]
\ead{fushuyue@buaa.edu.cn}
\credit{Data curation, Formal analysis, Methodology, Software, Writing - Original draft preparation, Writing - review $\&$ editing}
\affiliation[1]{organization={School of Astronautics, Beihang University},
                addressline={Xueyuan Road No.37}, 
                postcode={100191}, 
                city={Beijing},
                country={People's Republic of China}}

\affiliation[2]{organization={Shen Yuan Honors College, Beihang University},
                addressline={Xueyuan Road No.37}, 
                city={Beijing},
                postcode={100191}, 
                country={People's Republic of China}}

\affiliation[3]{organization={State Key Laboratory of High-Efficiency Reusable Aerospace Transportation Technology}, 
                postcode={102206}, 
                city={Beijing},
                country={People's Republic of China}}

\affiliation[4]{organization={Key Laboratory of Spacecraft Design Optimization $\&$ Dynamic Simulation Technologies, Ministry of Education}, 
                postcode={100191}, 
                city={Beijing},
                country={People's Republic of China}}

\author[1,4]{Di Wu}[style=chinese]
\ead{wudi2025@buaa.edu.cn}
\credit{Conceptualization, Funding acquisition, Writing - review $\&$ editing}

\author[1,2]{Yihan Peng}[style=chinese]
\ead{pengyihan@buaa.edu.cn}
\credit{Methodology, Formal analysis, Writing - review $\&$ editing}

\author[1,4]{Peng Shi}[style=chinese]
\ead{shipeng@buaa.edu.cn}
\credit{Methodology, Formal analysis, Writing - review $\&$ editing}

\author[1,3]{Shengping Gong}[style=chinese]
\cormark[1]
\ead{gongsp@buaa.edu.cn}
\credit{Conceptualization, Funding acquisition, Writing - review $\&$ editing}

\cortext[cor1]{Corresponding author}

\begin{abstract}
Distant prograde orbits around the Moon exhibit remarkable potential for practical applications such as cislunar surveillance activities and low-energy transfers due to their instability. Previous works on transfers from circular low Earth orbit to distant prograde orbits mainly focused on construction methods based on dynamical structures, lacking a comprehensive analysis of the solution space of this transfer scenario. This paper investigates the solution space and identifies families of transfers from a 167 km circular low Earth orbit to a 1:1 distant prograde orbit. In particular, grid search and trajectory continuation are performed to construct these transfer trajectories. Initial guesses of the transfers are selected in the 1:1 distant prograde orbit through a backward propagation strategy and are then corrected to satisfy specified constraints. Based on the obtained solutions, a linear predictor is derived to predict more feasible solutions and a predictor-corrector continuation method is used to extend the solution space. Twelve transfer families are identified, most of which are new or previously underexplored. The distributions of construction parameters and transfer characteristics of these twelve families are analyzed and discussed, showing which families are applicable to which types of specific practical missions. Comparison between the obtained solution and solution developed by previous works is further performed to imply the effects of the selection of dynamical model on transfer construction.
\end{abstract}

\begin{highlights}
\item Gird search and trajectory continuation are performed to explore the families of LEO-DPO transfers.
\item A new form of linear predictor is derived to continue the LEO-DPO transfer trajectories.
\item Families of LEO-DPO transfers are identified and analyzed.
\end{highlights}


\maketitle

\section{Introduction}\label{sec1}
An interest in Earth-Moon transfers has been renewed around the world as the the proposal and implementation of several missions, such as \textit{Artemis I} \citep{Batcha2020}, \textit{Chang’e-5} \citep{Zheng2023}, \textit{Chandrayaan-3} \citep{shetty2025rover}, and \textit{Danuri} \citep{Song2023}. The classical Earth-Moon transfer problem can be simplified into a scenario where the spacecraft departs from a circular low Earth orbit and then inserts into a circular low Moon orbit \cite{qi2016minimum,qi2016earth}. This scenario has been attracted considerable attention, leading to the development of several construction methods based on the patched two-body problem \cite{Battin1999}, multi-body problem \cite{Yagasaki2004a,Yagasaki2004b,topputo2013optimal,oshima2017analysis,oshima2019low}, and ephemeris model \cite{belbruno1993sun,campana2025ephemeris}. As the human's exploration of the Earth-Moon space continues, additional mission scenarios, including transfers to collinear libration points \cite{short2014lagrangian,Parker2014,oshima2025graph}, triangular libration points \cite{zhang2015transfer,tan2020single}, and distant retrograde orbits (DROs) \cite{demeyer2007transfer,capdevila2018transfer,zhang2021transfers,wang2025mechanism}, have been proposed to meet various practical requirements such as scientific observations and relay communication. Transfers to distant prograde orbits (DPOs) have also gained attention \cite{mingotti2012transfers}. The DPOs belong to $g$ and $g'$ families first discovered by H{\'e}non \cite{henon1969numerical} in the Hill problem, and later confirmed to exist in the planar circular restricted three-body problem (PCR3BP). When considering the DPOs around the Moon in the Earth-Moon PCR3BP, the instability of DPOs facilitates chaining with the L1/L2 collinear libration points \cite{parker2010chaining,gupta2021earth} and low-energy transfers \cite{mingotti2012transfers,gupta2021earth}, making them advantageous for cislunar surveillance activities and low-energy transfer missions. Therefore, investigating transfers to DPOs around the Moon is motivated by practical
mission requirements, such as cislunar surveillance. Since the DPOs around the Moon are periodic orbits in the Earth-Moon PCR3BP, we specifically focuses on bi-impulsive transfers from circular low Earth orbit (LEO) to DPOs in this dynamical model.

Bi-impulsive transfers refers to scenarios where the spacecraft departs from the Earth parking orbit (i.e., circular low Earth orbit) after perform an Earth injection impulse, coasts along the transfer trajectory, and finally inserts into the target Moon orbit (i.e., DPO) using a Moon insertion impulse. Consequently, these transfer trajectories can be considered as constrained trajectories in the multi-body problem, i.e., the states of trajectories should satisfy the constraints corresponding to the Earth parking orbit and target Moon orbit. When considering the construction problem of transfer trajectories in the multi-body problem, the numerical methods are required due to the absence of closed-form solutions in the PCR3BP \cite{Szebehely1967}. Typically, construction methods can be categorized into methods based on dynamical structures and methods based on grid search \cite{dutt2018review,scheuerle2025energy}. The methods based on dynamical structures use multi-body dynamical structures to analyze the motion characteristics of the spacecraft and then use these characteristics to construct transfer trajectories. For example, Belbruno and Miller \cite{belbruno1993sun} used the weak stability boundary to construct the low-energy transfers with lunar ballistic capture. Koon et al. \cite{Koon2001} and Dememeyer and Gurfil \cite{demeyer2007transfer} used invariant manifolds to construct transfers to Moon orbits and DROs, while Onozaki et al. \cite{Onozaki2017} and Fu et al. \cite{FU20254993,fu2025four} used Lagrangian coherent structures to design low-energy transfers with ballistic capture. For the scenarios transferring to the DPO around the Moon, Mingotti et al. \cite{mingotti2012transfers} calculated the stable manifold of the 1:1 DPO, and patched it with the unstable manifold in the Sun-Earth PCR3BP to generate initial guesses. Then, these initial guesses are corrected in the Sun-Earth/Moon planar bicircular restricted four-body problem (PBCR4BP). By using the invariant manifolds and the Sun-Earth/Moon PBCR4BP, they achieved single-impulse transfers (i.e., the value of the Moon insertion impulse is 0). Their work provided an effective framework to investigate the transfers to the DPO around the Moon. However, the construction methods based on dynamical structures only effectively generate low-energy transfers. They typically generate fewer solutions and lack global comprehensiveness. Alternatively, the methods based on grid search can be used. Topputo \cite{topputo2013optimal} and Oshima et al. \cite{oshima2019low} extensively investigated the solution space of transfers from the circular low Earth orbit to the circular low Moon orbit within 100 and 200 days, providing an useful perspective to understand the characteristic of this type of transfers. Methods based on grid search can effectively generate transfer families and provide a more comprehensive database for engineers to select suitable solutions satisfying various mission requirements. Moreover, to our best knowledge, this type of method specifically for the transfers from circular low Earth orbit to DPO and the corresponding transfer families remain inadequately explored. To obtain a more global map of solution space of transfers from circular low Earth orbit to DPO in the Earth-Moon PCR3BP and provide a more comprehensive database for the future practical missions, we use the methods based on grid search to construct the transfer trajectories and investigate the obtained families. 

Firstly, following Mingotti et al. \cite{mingotti2012transfers}, the 1:1 DPO is selected as the target Moon orbit. Then, we select the construction parameters corresponding to the states of the DPO, and use a backward propagation strategy \cite{capdevila2018transfer} to generate initial guesses. These initial guesses are corrected to satisfy constraints. Based on the obtained solutions, a predictor-corrector continuation method is used to further extend the solution space. In particular, the specific linear predictor suitable for this scenario is derived to predict more feasible solutions. As a result, we obtain $5663373$ solutions and identify 12 transfer families. Most of these 12 transfer families are new or underexplored compared to previous works on this transfer scenario \cite{mingotti2012transfers}. We perform a further analysis on these transfer families and present their distributions of construction parameters and characteristics. Finally, we present a discussion on which families are applicable to which types of specific practical missions and perform a comparison between our obtained solutions and solutions developed by previous works \cite{mingotti2012transfers}. Comparison suggests that using the Sun-Earth/Moon PBCR4BP may further reduce the total impulse of transfers, particularly the Moon insertion impulse.

The rest of this paper is organized as follows. Section \ref{sec2} presents the the mathematical background, including the Earth-Moon PCR3BP and DPOs. Section \ref{sec3} introduces the concept of bi-impulsive transfers and presents the methods to construct the transfers. Solutions obtained from these methods are presented in Section \ref{sec4} and the transfer families are identified and analyzed. Finally, conclusions are drawn in Section \ref{sec5}.

\section{Mathematical Background}\label{sec2}
In this section, the mathematical background of this study is presented, including the dynamical model of the Earth-Moon PCR3BP and the DPOs.
\subsection{Dynamical Model}\label{subsec2.1}
Since the DPOs are periodic orbits in the PCR3BP, to explore the transfers to the DPO around the Moon, the Earth-Moon PCR3BP is selected as the dynamical model in this paper. This model provides a higher fidelity than the two-body problem while maintaining lower complexity compared to the PBCR4BP and an ephemeris model \citep{mccarthy2021leveraging,FU2025}. Transfers constructed in this model are expected to involve direct transfers (Hohmann-like/Lambert-like solutions constructed from the two-body problem) and families of low-energy solutions. In this model, the Earth, the Moon, and spacecraft are assumed to move in the Earth-Moon orbital plane. The Earth and the Moon move in the circular orbit around their barycenter, and the spacecraft is treated as a massless point which does not affect the motion of the Earth and the Moon. When describing the dynamical equations of the Earth-Moon PCR3BP, the Earth-Moon rotating frame is adopted, as shown in Fig. \ref{EM_rot} \cite{FU20254993}. To improve the computational accuracy and simplify the expressions, we select the dimensionless units as follows: the length unit (LU) is set as the Earth-Moon distance, the mass unit (MU) is set as the combined mass of the Earth and Moon, and the time unit (TU) is set as ${\text{TU}}={T_\text{EM}}/{2\pi}$, where ${T_{{\text{EM}}}}$ is the orbital period of the Earth and Moon around their barycenter. Then, the dynamical equations of the Earth-Moon PCR3BP are expressed as:

\begin{figure}[h]
\centering
\includegraphics[width=0.42\textwidth]{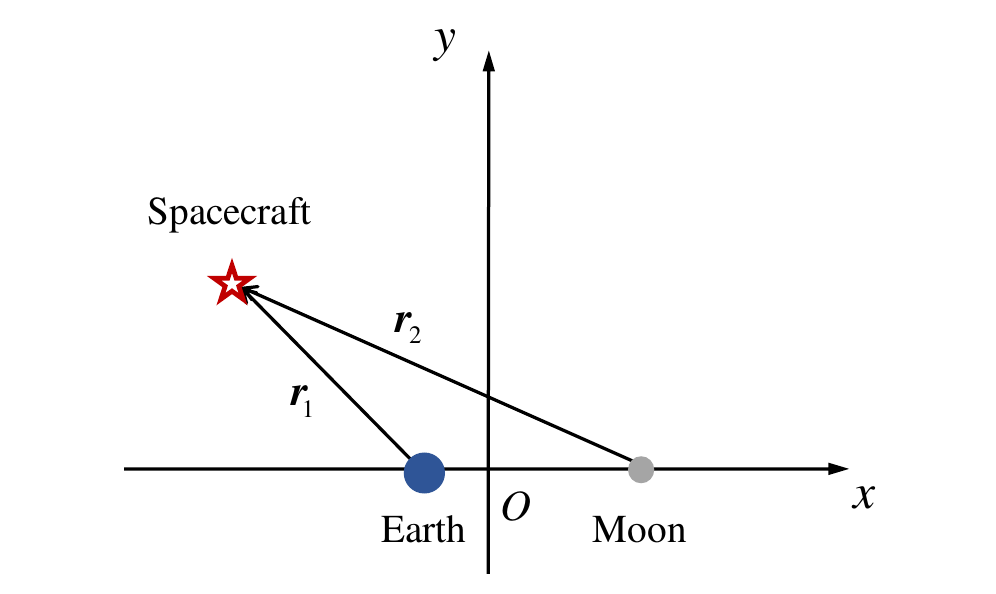}
\caption{Schematic of the Earth-Moon rotating frame.}
\label{EM_rot}
\end{figure}

\begin{equation}
\left[ {\begin{array}{*{20}{c}}
{\begin{array}{*{20}{c}}
{\dot x}\\
{\dot y}
\end{array}}\\
{\begin{array}{*{20}{c}}
{\dot u}\\
{\dot v}
\end{array}}
\end{array}} \right] = \left[ {\begin{array}{*{20}{c}}
{\begin{array}{*{20}{c}}
u\\
v
\end{array}}\\
{\begin{array}{*{20}{c}}
{2v + \frac{{\partial {\Omega _3}}}{{\partial x}}}\\
{ - 2u + \frac{{\partial {\Omega _3}}}{{\partial y}}}
\end{array}}
\end{array}} \right]\label{eq1}
\end{equation}

\begin{equation}
{\Omega _3} = \frac{1}{2}\left[ {{x^2} + {y^2} + \mu \left( {1 - \mu } \right)} \right] + \frac{{1 - \mu }}{{{r_1}}} + \frac{\mu }{{{r_2}}}\label{eq2}
\end{equation}
where $\bm{X} = \left[ x, y, u, v\right]^{\text{T}}$ denotes the orbital state in the Earth-Moon rotating frame, and ${\Omega _3}$ is the effective potential of the Earth-Moon PCR3BP. The parameter $\mu $ denotes the Earth-Moon mass parameter which is calculated by $\mu=m_{\text{M}}/\left(m_{\text{E}}+m_{\text{M}}\right)$,where $m_{\text{E}}$ and $m_{\text{M}}$ denote the masses of the Earth and the Moon. The parameters $r_1$ and $r_2$ denote the distances between the spacecraft and the Earth ($r_1$) and the Moon ($r_2$), which are expressed as:
\begin{equation}
{r_1} = \sqrt {{{\left( {x + \mu } \right)}^2} + {y^2}} \text{ }\text{ }\text{ }\text{ }{r_2} = \sqrt {{{\left( {x + \mu - 1} \right)}^2} + {y^2}}\label{eq3}
\end{equation}
For a specific trajectory in the Earth-Moon PCR3BP, there exists a constant, namely, the Jacobi energy. This constant is expressed as \cite{FU2025}:
\begin{equation}
C = -\left( {{u^2} + {v^2}} \right) +  \left( {{x^2} + {y^2}} \right) + \frac{{2(1 - \mu) }}{{{r_1}}} + \frac{2\mu }{{{r_2}}} + \mu \left( {1 - \mu } \right) \label{eq4}
\end{equation}
Since these are not closed-form solutions in the PCR3BP, the numerical methods are used to propagate trajectories. We use the variable step-size, variable order (VSVO) Adams-Bashforth-Moulton algorithm with absolute and relative tolerances set to $1 \times 10^{-13}$, achieved by MATLAB®'s ode113 command \cite{oshima2021capture}. The specific values of the parameters used are presented in Table \ref{tab1} \cite{FU2025}.
\begin{table}[!htb]
\caption{Parameter Setting for the Earth-Moon PCR3BP}\label{tab1}%
\centering
\renewcommand{\arraystretch}{1.5}
\begin{tabular}{@{}llll@{}}
\hline
Symbol & Value  & Units & Meaning\\
\hline
$\mu$    & $1.21506683 \times {10^{ - 2}}$   & --  & Earth-Moon mass parameter  \\
${T_{{\text{EM}}}}$    & $2.24735067 \times {10^6}$   & s  & Earth-Moon period  \\
$R_{\text{E}}$    & $6378.145$   & km  & Mean Earth’s radius  \\
$R_{\text{M}}$    & $1737.100$   & km  & Mean Moon’s radius  \\
LU    & $3.84405 \times {10^5}$   & km  & Length unit  \\
TU    & $3.7567696752 \times {10^5}$   & s  & Time unit  \\
\hline
\end{tabular}
\end{table}

Subsequently, the background about the DPO around the Moon in the Earth-Moon PCR3BP is introduced.
\subsection{DPO}\label{subsec2.2}
\begin{figure}[h]
\centering
\includegraphics[width=0.6\textwidth]{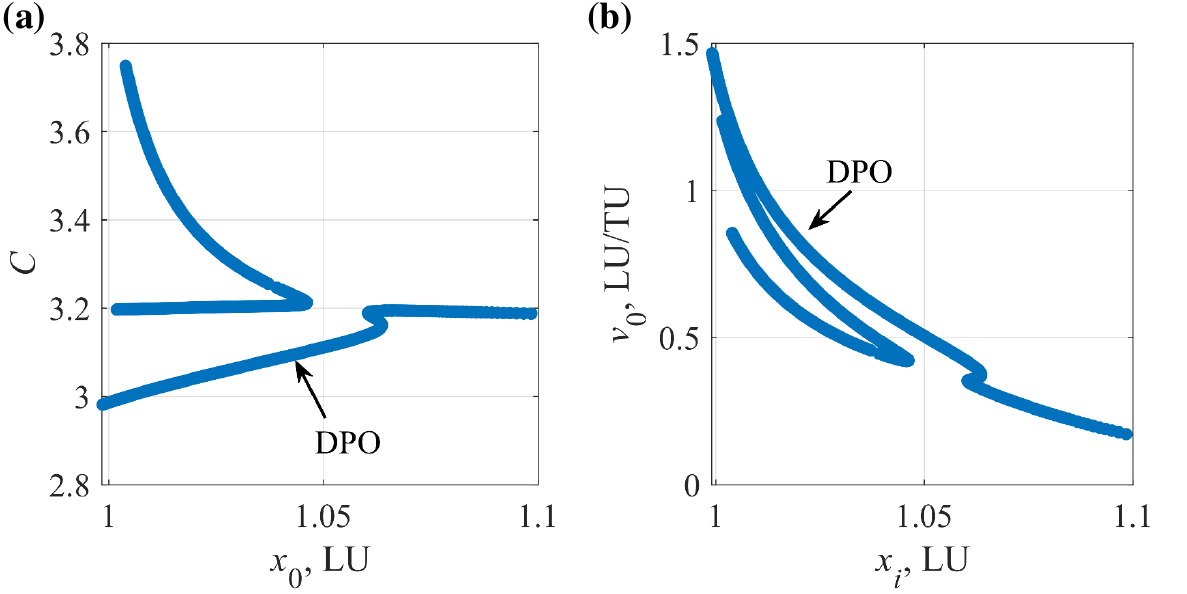}
\caption{The distribution of $g$ and $g'$ families in the Earth-Moon PCR3BP. (a) Distribution in terms of $\left({x_0},\text{ }{C}\right)$; (b) Distribution in terms of $\left({x_0},\text{ }{v_0}\right)$.}
\label{fig_DPO_g}
\end{figure}

The DPOs, first discovered by H{\'e}non \cite{henon1969numerical} in the Hill problem, belong to the $g$ and $g'$ families of the periodic orbits. The $g$ and $g'$ families exist in both the Hill problem and the PCR3BP \cite{dei2018survey}. These axisymmetric orbits around the secondary body satisfy the following constraints:
\begin{equation}
\bm{c}\left( \bm{X }_0 \right) = \phi _{{t_0}}^{{t_0} + T}\left( {{\bm{X}_0}} \right)-{\bm{X }_0} = \textbf{0}
\label{eq5}
\end{equation}
where $\phi _{{t_0}}^{{t_0} + T}:\mathbb{R} \times \mathbb{R} \times {\text{D}} \to {\text{D;   }}\left( {{t_0},{\text{ }}{t_0} + T,{\text{ }}{\bm{X}_0}} \right) \to \phi _{{t_0}}^{{t_0} + T}\left( {{\bm{X}_0}} \right)$ denotes the dynamical flow of the Earth-Moon PCR3BP, $\bm{X}_0$ denotes the initial states of the DPO, and $T$ denotes the period of the DPO. In this paper, $\bm{X}_0$ is expressed as $\bm{X}_0=\left[x_0,\text{ }0,\text{ }0,\text{ }v_0\right]^\text{T}$. The corresponding distribution of initial states of these periodic orbits in the Earth-Moon PCR3BP is shown in Fig. \ref{fig_DPO_g}. The DPOs are labeled in the figure. Unlike the bifurcation phenomenon reported in Ref. \cite{henon1969numerical} in the Hill problem, the distribution shown in Fig. \ref{fig_DPO_g} exhibits that the orbit families are divided into two branches, which is similar to the distribution in the Sun-Mars PCR3BP reported by Dei Tos et al. \cite{dei2018survey}.

\begin{figure}[h]
\centering
\includegraphics[width=0.35\textwidth]{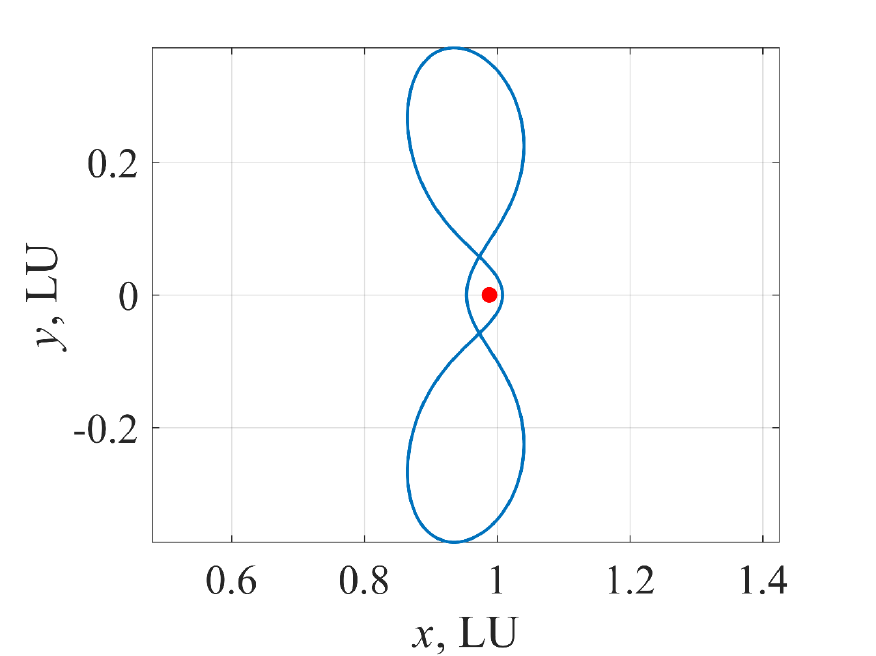}
\caption{The selected DPO. The red dot denotes the Moon.}
\label{DPO}
\end{figure}

For DPOs, there exists a specific range of the Jacobi energy where the DPOs are unstable orbits measured by the linear stability indices \cite{mingotti2012transfers,gupta2021earth}. This instability facilitates chaining DPO with the L1/L2 libration points, i.e., natural homoclinic or heteroclinic chains \cite{parker2010chaining,gupta2021earth,du2022transfer}. These chains provide a natural pathway for constructing transfers between libration points and transfers to resonant orbits facilitating cislunar surveillance activities \cite{gupta2021earth,du2022transfer,du2023two,jiao2024asteroid}. Moreover, the DPO itself facilitates low-energy Earth-Moon transfers due to its instability \cite{mingotti2012transfers}. Considering the remarkable potential of the DPO in practical applications, we select the DPO as the target Moon orbit in this paper. Previous work by Mingotti et al. \cite{mingotti2012transfers} identified few solutions of transfers without investigating transfer families. Therefore, following the method adopted by Topputo \cite{topputo2013optimal}, Oshima et al. \cite{oshima2019low}, and Capdevila and Howell \cite{capdevila2018transfer} in the transfers from LEO to low lunar orbits and DROs, we perform a grid search and trajectory continuation of LEO-DPO transfers and explore the transfer families existing in the Earth-Moon PCR3BP. The selected DPO is a 1:1 DPO, i.e., the period of the DPO is $2\pi \text{  TU}$. The selected DPO is shown in Fig. \ref{DPO}, and its initial states are presented in Table \ref{DPO_initial_state}. This DPO is an unstable orbit, and we focus on the bi-impulsive transfers this DPO. The concept and construction of this scenario are detailed in Section \ref{sec3}.

\begin{table}
\caption{Initial States of The Selected DPO}\label{DPO_initial_state}%
\centering
\renewcommand{\arraystretch}{1.5}
\begin{tabular}{@{}lll@{}}
\hline
Parameter & Value  & Unit \\
\hline
$x_{0}$ & $1.007819412874657$ & LU \\
$v_{0}$ & $1.082615000979063$ & LU/TU \\
\hline
\end{tabular}
\end{table}

\section{LEO-DPO Transfer}\label{sec3}
In this section, the concept of bi-impulsive LEO-DPO transfer is introduced. Then, the construction method of this type of transfer is presented, including performing grid search and trajectory continuation.
\subsection{Concept of Bi-Impulsive LEO-DPO Transfer}\label{subsec3.1}
The schematic of bi-impulsive LEO-DPO transfer in the Earth-Moon rotating frame is presented in Fig. \ref{LEO_DPO_Transfer}. This type of transfer describes a scenario where the spacecraft departs from a LEO by an Earth injection impulse ($\Delta{v}_i$), coasts along the transfer trajectory, and finally inserts into the selected DPO after performing an insertion impulse ($\Delta{v}_f$), where the subscripts ‘\textit{i}’ and ‘\textit{f}’ denote quantities corresponding with the departure and the insertion points, respectively. Similar to other types of Earth-Moon transfers \cite{topputo2013optimal,oshima2019low,capdevila2018transfer}, the impulses are assumed tangential. Therefore, the constraints of bi-impulsive LEO-DPO transfer can be expressed as:

\begin{figure}[h]
\centering
\includegraphics[width=0.35\textwidth]{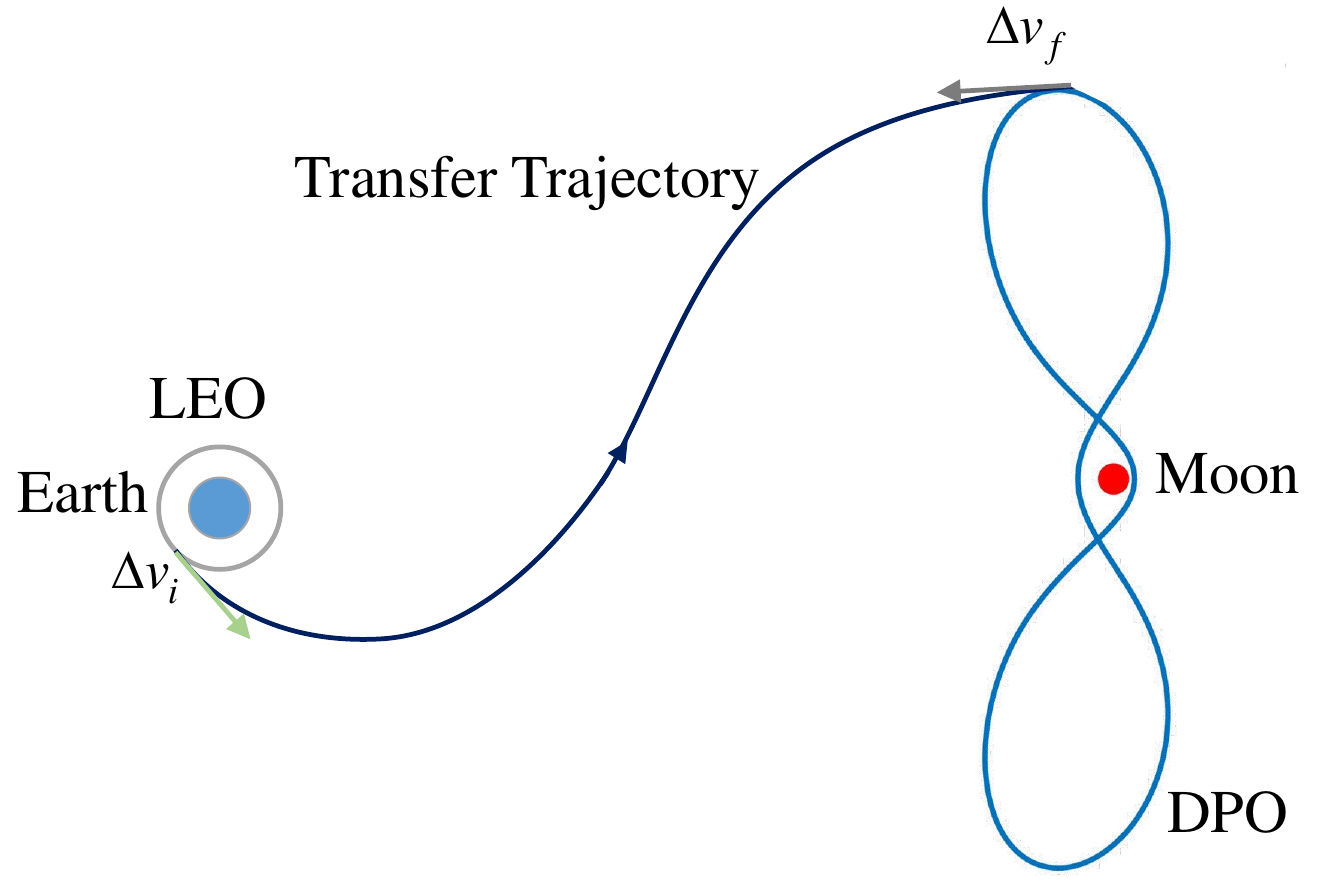}
\caption{Schematic of bi-impulsive LEO-DPO transfer in the Earth-Moon rotating frame.}
\label{LEO_DPO_Transfer}
\end{figure}

\begin{equation}
{\bm{\psi }_i} = \left[ {\begin{array}{*{20}{c}}
  {{{\left( {{x_i} + \mu } \right)}^2} + {y_i}^2 - {{\left( {{R_{\text{E}}} + {h_i}} \right)}^2}} \\ 
  {\left( {{x_i} + \mu } \right)\left( {{u_i} - {y_i}} \right) + {y_i}\left( {{v_i} + {x_i} + \mu } \right)} 
\end{array}} \right] = \mathbf{0} \label{eq6}
\end{equation}

\begin{equation}
{\bm{\psi }_f} = \left[ {\begin{array}{*{20}{c}}
  x_f-x_{\text{DPO}} \\ 
   y_f-y_{\text{DPO}}\\
  u_f v _{\text{DPO}}-v_f u _{\text{DPO}}
\end{array}} \right] = \mathbf{0} \label{eq7}
\end{equation}
where $h_i$ denotes the orbital altitude of the LEO and the subscript ‘DPO’ denotes the orbital states of the DPO. To compare the obtained solutions with those obtained by Ref. \cite{mingotti2012transfers}, $h_i$ is set to 167 km. Satisfying these constraints, the impulses can be calculated by:
\begin{equation}
\Delta {v_i} = \sqrt {{{\left( {{u_i} - {y_i}} \right)}^2} + {{\left( {{v_i} + {x_i} + \mu } \right)}^2}}  - \sqrt {\frac{{1 - \mu }}{{{R_{\text{E}}} + {h_i}}}}  \label{eq8}
\end{equation}
\begin{equation}
\Delta {v_f} = \sqrt {\left({{u_f}}-{u_\text{DPO}}\right)^2 + \left({{v_f}}-{v_\text{DPO}}\right)^2} \label{eq9}
\end{equation}
\begin{equation}
\Delta {v} = \Delta {v_i} + \Delta {v_f} \label{eq10}
\end{equation}
where $\Delta {v}$ denotes the total impulse. Based on the aforementioned discussion, we present the grid search and trajectory continuation methods to construct the bi-impulsive LEO-DPO transfers.
\subsection{Construction Method}\label{subsec3.2}
The construction method is presented in this subsection. To ensure the tangential constraints corresponding to the insertion point, we select the construction parameters at the insertion point and perform a backward strategy \cite{capdevila2018transfer} to construct the transfers. The construction parameters are set as:
\begin{equation}
\bm{y} = {\left[ {{\tau _f},\text{ }{\beta _f},\text{ }{\text{TOF}}} \right]^{\text{T}}} \label{eq11}
\end{equation}
where $\tau _f$ denotes the time phase in the DPO ($0\leq \tau _f \leq 2\pi\text{ } \left(\text{TU}\right)$), $\beta _f$ denotes the velocity ratio at the insertion point, and ${\text{TOF}}$ denotes the time of flight (TOF). With this setting, the states at the insertion point can be expressed as (shown in Fig. \ref{construction_parameter}):

\begin{figure}[h]
\centering
\includegraphics[width=0.27\textwidth]{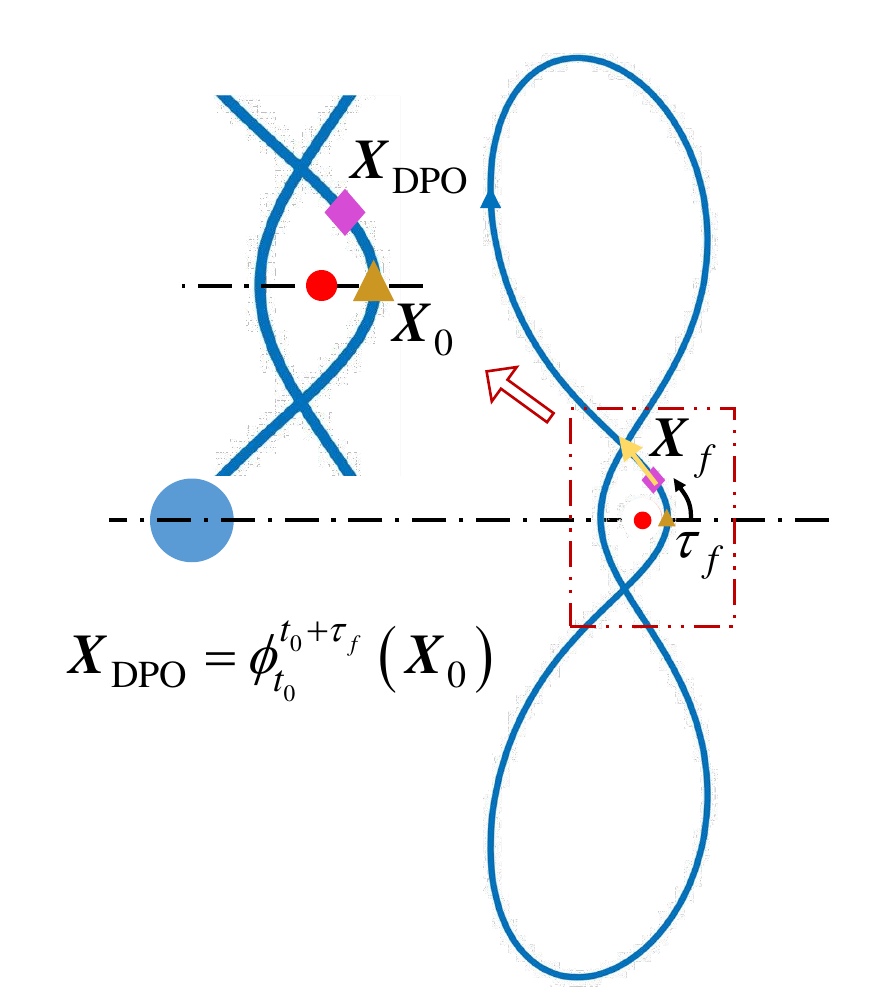}
\caption{Schematic of construction parameters at the insertion point.}
\label{construction_parameter}
\end{figure}

\begin{equation}
\left\{ \begin{gathered}
  {\bm{X}_{{\text{DPO}}}} = \phi _{{t_0}}^{{t_0} + {\tau _f}}\left( {{\bm{X}_0}} \right) \hfill \\
  {x_f} = {x_{{\text{DPO}}}} \hfill \\
  {y_f} = {y_{{\text{DPO}}}} \hfill \\
  {u_f} = {\beta _f}{u_{{\text{DPO}}}} \hfill \\
  {v_f} = {\beta _f}{v_{{\text{DPO}}}} \hfill \\ 
\end{gathered}  \right.
 \label{eq12}
\end{equation}
Generally, $t_0$ is set to 0. Obtaining the states of the insertion point, we perform a \textbf{backward-in-time propagation} to search the transfer trajectories satisfying the constraints. The detailed steps are presented in the following texts, including constructing initial guesses, performing correction, and performing trajectory continuation.
\subsubsection{Performing Grid Search}\label{subsubsec3.2.1}
Based on aforementioned the setting of construction parameters, we set time phase in the DPO as $\tau_f \in \left[0,\text{ }2\pi\right]$ with a step-size of $\pi/5000$, and set the velocity ratio at the insertion point as $\beta_f \in \left[1,\text{ }2\right] $ with a step-size of $0.0001$. After calculating the states at the insertion point according to Eq. \eqref{eq12}, we perform a \textbf{backward-in-time propagation}, and the propagation time is set to $12\pi$. Since the selected construction parameters satisfy $\bm{\psi}_f$ rigorously, when the states during the propagation satisfy \cite{fu2025constructing}:
\begin{equation}
\left\| {{\bm{\psi} _i}} \right\| < 1 \times {10^{ - 4}}
\label{eq13}
\end{equation}
the corresponding $\tau_f$ and $\beta_f$ are recorded as the initial guesses of the states at the insertion point, and the corresponding propagation time is recorded as an initial guess of $\text{TOF}$ (notably, $0<\text{TOF} \leq 12\pi$).
When obtaining the initial guesses, trajectory correction is performed to make the trajectories satisfy the constraints $\bm{\psi}_i$. The correction is achieved by MATLAB®'s fsolve command, using the Levenberg-Marquardt method. The function is set as $\bm{\psi}_i$. The parameter settings for fsolve command is summarized in Table \ref{tab4} (the other parameters are set to default values). After correction, we select the solutions with $\left\| {{\bm{\psi} _i}} \right\| < 5 \times {10^{ - 8}}$, this tolerance ensures that the error of the $h_i$ is less than 1 km. 

\begin{table}[!htb]
\caption{Parameter settings for fsolve command}\label{tab4}%
\centering
\renewcommand{\arraystretch}{1.5}
\begin{tabular}{@{}ll@{}}
\hline
Parameter & Value   \\
\hline
TolX    & $1 \times {10^{ - 16}}$       \\
TolFun    & $1 \times {10^{ - 16}}$       \\
MaxIter    & $800$    \\
\hline
\end{tabular}
\end{table}

\subsubsection{Performing Trajectory Continuation}\label{subsubsec3.2.2}
After obtaining solutions from method mentioned in Section \ref{subsubsec3.2.1}, we perform a trajectory continuation based on the obtained solutions to search more feasible solutions and explore the transfer families \cite{topputo2013optimal,oshima2019low,capdevila2018transfer}. Generally, the continuation method is a predictor-corrector method \cite{fu2025analytical}. When obtaining a feasible solution satisfying constraints $\bm{\psi}_i$, the feasible direction of the next feasible solution near the existing feasible solution is predicted by several methods, and the initial guess of the next feasible solution is obtained. Correction is then performed to make the trajectory satisfying constraints. Since the trajectory correction is introduced in Section \ref{subsubsec3.2.1}, in this subsection, we focus on the predictor. To predict the feasible direction accurately and conveniently, several predictors have been proposed, such as natural parameter predictor \cite{topputo2013optimal,oshima2019low,singh2019mission}, linear predictor \cite{capdevila2018transfer,pushparaj2021transfers,singh2021exploiting,pushparaj2024optimization}, and nonlinear predictor \cite{fu2025analytical}. Among them, the natural parameter predictor only make a minor adjustment on the existing solution (denoted as $\bm{y}^0$), while linear and nonlinear predictors predict the feasible direction based on Taylor expansion of the constraints with respect to $\bm{y}^0$. Based on the aforementioned discussion, the linear predictor improve the prediction accuracy compared to the natural parameter predictor, while maintaining lower complexity compared to the nonlinear predictor. Therefore, we adopt the linear predictor to perform the prediction step of trajectory continuation. Considering the computational efficiency, we only select the solutions with $\Delta{v}<3.45\text{ km/s}$ obtained from the method mentioned in the Section \ref{subsubsec3.2.1} and perform trajectory continuation. With the linear approximation of the constraints $\bm{\psi}_i$, we obtain the relationship between $\bm{\psi}_i\left(\bm{y}^1\right)$ and $\bm{\psi}_i\left(\bm{y}^0\right)$ ($\bm{y}^1$ is the next feasible solution near the existing feasible solution $\bm{y}^0$):
\begin{equation}
{\bm{\psi }_i}\left( {{\bm{y}^1}} \right) - {\bm{\psi }_i}\left( {{\bm{y}^0}} \right) \approx \left.\frac{{\partial {\bm{\psi }_i}}}{{\partial \bm{y}}}\right|_{{\bm{y} = {\bm{y}^0}}}\left( {{\bm{y}^1} - {\bm{y}^0}} \right) = \left.\frac{{\partial {\bm{\psi }_i}}}{{\partial \bm{y}}}\right|_{{\bm{y} = {\bm{y}^0}}}\delta \bm{y}\Delta s = \bm{0}
\label{eq14}
\end{equation}
where $\delta \bm{y}$ denotes the feasible direction ($\left|\left|\delta \bm{y}\right|\right|=1$), and $\Delta s$ denotes the step-size in the continuation method. Equation \eqref{eq14} can be further simplified into:
\begin{equation}
\left.\frac{{\partial {\bm{\psi }_i}}}{{\partial \bm{y}}}\right|_{{\bm{y} = {\bm{y}^0}}}\delta \bm{y} = \bm{A}\delta \bm{y} =\bm{0}
\label{eq15}
\end{equation}
which is a linear equation system, where the matrix $\bm{A}$ is a $2\times3$ matrix. The components of the matrix $\bm{A}$ can be expressed as ($\bm{\psi}_i=\left[\psi_{i1},\text{ }\psi_{i2}\right]^{\text{T}}$):
\begin{equation}
\begin{gathered}
  {\bm{A}_{11}} = \frac{{\partial {\psi _{i1}}}}{{\partial {\tau _f}}} = 2\left( {{x_i} + \mu } \right)\frac{{\partial {x_i}}}{{\partial {\tau _f}}} + 2{y_i}\frac{{\partial {y_i}}}{{\partial {\tau _f}}} \hfill \\
  {\bm{A}_{12}} = \frac{{\partial {\psi _{i1}}}}{{\partial {\beta _f}}} = 2\left( {{x_i} + \mu } \right)\frac{{\partial {x_i}}}{{\partial {\beta _f}}} + 2{y_i}\frac{{\partial {y_i}}}{{\partial {\beta _f}}} \hfill \\
  {\bm{A}_{13}} = \frac{{\partial {\psi _{i1}}}}{{\partial {\text{TOF}}}} = 2\left( {{x_i} + \mu } \right)\frac{{\partial {x_i}}}{{\partial {\text{TOF}}}} + 2{y_i}\frac{{\partial {y_i}}}{{\partial {\text{TOF}}}} \hfill \\ 
\end{gathered}
\label{eq16}
\end{equation}
\begin{equation}
\begin{gathered}
  {\bm{A}_{21}} = \frac{{\partial {\psi _{i2}}}}{{\partial {\tau _f}}} = \frac{{\partial {x_i}}}{{\partial {\tau _f}}}\left( {{u_i} - {y_i}} \right) + \left( {{x_i} + \mu } \right)\left( {\frac{{\partial {u_i}}}{{\partial {\tau _f}}} - \frac{{\partial {y_i}}}{{\partial {\tau _f}}}} \right) + \frac{{\partial {y_i}}}{{\partial {\tau _f}}}\left( {{v_i} + {x_i} + \mu } \right) + {y_i}\left( {\frac{{\partial {v_i}}}{{\partial {\tau _f}}} + \frac{{\partial {x_i}}}{{\partial {\tau _f}}}} \right) \hfill \\
  {\bm{A}_{22}} = \frac{{\partial {\psi _{i2}}}}{{\partial {\beta _f}}} = \frac{{\partial {x_i}}}{{\partial {\beta _f}}}\left( {{u_i} - {y_i}} \right) + \left( {{x_i} + \mu } \right)\left( {\frac{{\partial {u_i}}}{{\partial {\beta _f}}} - \frac{{\partial {y_i}}}{{\partial {\beta _f}}}} \right) + \frac{{\partial {y_i}}}{{\partial {\beta _f}}}\left( {{v_i} + {x_i} + \mu } \right) + {y_i}\left( {\frac{{\partial {v_i}}}{{\partial {\beta _f}}} + \frac{{\partial {x_i}}}{{\partial {\beta _f}}}} \right) \hfill \\
  {\bm{A}_{23}} = \frac{{\partial {\psi _{i2}}}}{{\partial {\text{TOF}}}} = \frac{{\partial {x_i}}}{{\partial {\text{TOF}}}}\left( {{u_i} - {y_i}} \right) + \left( {{x_i} + \mu } \right)\left( {\frac{{\partial {u_i}}}{{\partial {\text{TOF}}}} - \frac{{\partial {y_i}}}{{\partial {\text{TOF}}}}} \right) + \frac{{\partial {y_i}}}{{\partial {\beta _f}}}\left( {{v_i} + {x_i} + \mu } \right) + {y_i}\left( {\frac{{\partial {v_i}}}{{\partial {\text{TOF}}}} + \frac{{\partial {x_i}}}{{\partial {\text{TOF}}}}} \right) \hfill \\ 
\end{gathered}
\label{eq18}
\end{equation}
where
\begin{align}\label{eq188}
  \frac{{\partial {x_i}}}{{\partial {\tau _f}}} &= \frac{{\partial {x_i}}}{{\partial {x_f}}}\frac{{\partial {x_f}}}{{\partial {\tau _f}}} + \frac{{\partial {x_i}}}{{\partial {y_f}}}\frac{{\partial {y_f}}}{{\partial {\tau _f}}} + \frac{{\partial {x_i}}}{{\partial {u_f}}}\frac{{\partial {u_f}}}{{\partial {\tau _f}}} + \frac{{\partial {x_i}}}{{\partial {v_f}}}\frac{{\partial {v_f}}}{{\partial {\tau _f}}} \\ 
   \notag&= \frac{{\partial {x_i}}}{{\partial {x_f}}}\frac{{\partial {x_f}}}{{\partial {x_{{\text{DPO}}}}}}\frac{{\partial {x_{{\text{DPO}}}}}}{{\partial {\tau _f}}} + \frac{{\partial {x_i}}}{{\partial {y_f}}}\frac{{\partial {y_f}}}{{\partial {y_{{\text{DPO}}}}}}\frac{{\partial {y_{{\text{DPO}}}}}}{{\partial {\tau _f}}} + \frac{{\partial {x_i}}}{{\partial {u_f}}}\frac{{\partial {u_f}}}{{\partial {u_{{\text{DPO}}}}}}\frac{{\partial {u_{{\text{DPO}}}}}}{{\partial {\tau _f}}} + \frac{{\partial {x_i}}}{{\partial {v_f}}}\frac{{\partial {v_f}}}{{\partial {v_{{\text{DPO}}}}}}\frac{{\partial {v_{{\text{DPO}}}}}}{{\partial {\tau _f}}} \\ 
  \notag& = \frac{{\partial {x_i}}}{{\partial {x_f}}}{{\dot x}_{{\text{DPO}}}} + \frac{{\partial {x_i}}}{{\partial {y_f}}}{{\dot y}_{{\text{DPO}}}} + \frac{{\partial {x_i}}}{{\partial {u_f}}}{\beta _f}{{\dot u}_{{\text{DPO}}}} + \frac{{\partial {x_i}}}{{\partial {v_f}}}{\beta _f}{{\dot v}_{{\text{DPO}}}} 
\end{align}
\begin{equation}
\frac{{\partial {y_i}}}{{\partial {\tau _f}}} = \frac{{\partial {y_i}}}{{\partial {x_f}}}{\dot x_{{\text{DPO}}}} + \frac{{\partial {y_i}}}{{\partial {y_f}}}{\dot y_{{\text{DPO}}}} + \frac{{\partial {y_i}}}{{\partial {u_f}}}{\beta _f}{\dot u_{{\text{DPO}}}} + \frac{{\partial {y_i}}}{{\partial {v_f}}}{\beta _f}{\dot v_{{\text{DPO}}}}
\label{eq19}
\end{equation}
\begin{equation}
\frac{{\partial {u_i}}}{{\partial {\tau _f}}} = \frac{{\partial {u_i}}}{{\partial {x_f}}}{\dot x_{{\text{DPO}}}} + \frac{{\partial {u_i}}}{{\partial {y_f}}}{\dot y_{{\text{DPO}}}} + \frac{{\partial {u_i}}}{{\partial {u_f}}}{\beta _f}{\dot u_{{\text{DPO}}}} + \frac{{\partial {u_i}}}{{\partial {v_f}}}{\beta _f}{\dot v_{{\text{DPO}}}}
\label{eq20}
\end{equation}
\begin{equation}
\frac{{\partial {v_i}}}{{\partial {\tau _f}}} = \frac{{\partial {v_i}}}{{\partial {x_f}}}{\dot x_{{\text{DPO}}}} + \frac{{\partial {v_i}}}{{\partial {y_f}}}{\dot y_{{\text{DPO}}}} + \frac{{\partial {v_i}}}{{\partial {u_f}}}{\beta _f}{\dot u_{{\text{DPO}}}} + \frac{{\partial {v_i}}}{{\partial {v_f}}}{\beta _f}{\dot v_{{\text{DPO}}}}
\label{eq21}
\end{equation}
\begin{align}
\frac{{\partial {x_i}}}{{\partial {\beta _f}}} &= \frac{{\partial {x_i}}}{{\partial {x_f}}}\frac{{\partial {x_f}}}{{\partial {\beta _f}}} + \frac{{\partial {x_i}}}{{\partial {y_f}}}\frac{{\partial {y_f}}}{{\partial {\beta _f}}} + \frac{{\partial {x_i}}}{{\partial {u_f}}}\frac{{\partial {u_f}}}{{\partial {\beta _f}}} + \frac{{\partial {x_i}}}{{\partial {v_f}}}\frac{{\partial {v_f}}}{{\partial {\beta _f}}} \hfill \\
   \notag&= \frac{{\partial {x_i}}}{{\partial {u_f}}}{u_{{\text{DPO}}}} + \frac{{\partial {x_i}}}{{\partial {v_f}}}{v_{{\text{DPO}}}}
\label{eq22}
\end{align}
\begin{equation}
\frac{{\partial {y_i}}}{{\partial {\beta _f}}} = \frac{{\partial {y_i}}}{{\partial {u_f}}}{u_{{\text{DPO}}}} + \frac{{\partial {y_i}}}{{\partial {v_f}}}{v_{{\text{DPO}}}}
\label{eq23}
\end{equation}
\begin{equation}
\frac{{\partial {u_i}}}{{\partial {\beta _f}}} = \frac{{\partial {u_i}}}{{\partial {u_f}}}{u_{{\text{DPO}}}} + \frac{{\partial {u_i}}}{{\partial {v_f}}}{v_{{\text{DPO}}}}
\label{eq24}
\end{equation}
\begin{equation}
\frac{{\partial {v_i}}}{{\partial {\beta _f}}} = \frac{{\partial {v_i}}}{{\partial {u_f}}}{u_{{\text{DPO}}}} + \frac{{\partial {v_i}}}{{\partial {v_f}}}{v_{{\text{DPO}}}}
\label{eq25}
\end{equation}
\begin{equation}
\frac{{\partial {x_i}}}{{\partial {\text{TOF}}}} =  - {\dot x_i}
\label{eq26}
\end{equation}
\begin{equation}
\frac{{\partial {y_i}}}{{\partial {\text{TOF}}}} =  - {\dot y_i}
\label{eq27}
\end{equation}
\begin{equation}
\frac{{\partial {u_i}}}{{\partial {\text{TOF}}}} =  - {\dot u_i}
\label{eq28}
\end{equation}
\begin{equation}
\frac{{\partial {v_i}}}{{\partial {\text{TOF}}}} =  - {\dot v_i}
\label{eq29}
\end{equation}
where $\partial {\bm{X}_i}/\partial {\bm{X}_f}=\bm{\Phi}\left(t_i,\text{ }t_f\right)$ denotes the state transition matrix of the transfer trajectory from $t_f$ to $t_i$, which can calculated by:
\begin{equation}
\left\{ \begin{gathered}
  {{\bm{\dot \Phi }}} = \frac{{\partial \bm{f}}}{{\partial \bm{X}}}{\bm{\Phi }} \hfill \\
  {\bm{\Phi }}\left(t_f,\text{ }t_f\right) = {\bm{I}_{6 \times 6}} \hfill \\ 
\end{gathered}  \right.\label{eq30}
\end{equation}
Similar to the construction of the transfer trajectories, Eq. \eqref{eq30} is also calculated by the back-in-time propagation. When obtaining the matrix $\bm{A}$, to obtain $\delta \bm{y}$, the singular value decomposition of the matrix $\bm{A}=\bm{A}_{2 \times 3}$ can be expressed as:
\begin{equation}
\bm{A} = \bm{U}_{2 \times 2}  \bm{\Sigma }_{2 \times 3} \bm{V}_{3 \times 3}^{{\text{T}}}
\label{eq31}
\end{equation}
where the matrices $\bm{U}_{2 \times 2}$ and $\bm{V}_{3 \times 3}$ are orthogonal matrices. The matrix $\bm{\Sigma}_{2 \times 3}$ is the singular value matrix of the matrix $\bm{A}$. The feasible direction $\delta \bm{y}$ is obtained from the least squares method ($\bm{V}_{3 \times 3}=\left[\bm{V}_1,\text{ }\bm{V}_2,\text{ }\bm{V}_3\right]$):
\begin{equation}
\delta \bm{y} = \bm{V}_3
\label{eq32}
\end{equation}
When obtaining the feasible direction, we construct the initial guess ($\Delta s=1\times10^{-5} \text{ and } -1\times10^{-5}$):
\begin{equation}
{\bm{\tilde y}^1} = {\bm{y}^0} + \delta \bm{y}\Delta s
\label{eq33}
\end{equation}
and correct it to satisfy the constraints $\bm{\psi}_i$. When the new feasible solution is obtained, the aforementioned procedure continues until the termination criteria are satisfied as follows:

\begin{enumerate}[label=(\arabic*)]
\item The tolerance of the constraints $\bm{\psi}_i$ after differential correction exceeds the preset tolerance, i.e., $\left\| \bm{\psi}_i \right\| > 5\times 10^{-8}$;
\item The number of predictor-corrector steps ($j$) exceeds the maximum number of predictor-corrector steps, i.e., $j>100000$.
\end{enumerate}
Summarizing the solutions obtained in Sections \ref{subsubsec3.2.1}-\ref{subsubsec3.2.2}, we then present the transfer families found in this paper.
\section{Results And Discussion}\label{sec4}
In this section, the solutions obtained from the methods described in Section \ref{sec3} are presented and discussed. Specifically, we identify 12 transfer families, most of which are new or less-known families existing in the previous works \cite{mingotti2012transfers}. Then, we discuss on the distribution of the construction parameters and characteristics of the obtained transfer families.
\subsection{Overview of Obtained Solutions}\label{subsec4.1}
\begin{figure}[h]
\centering
\includegraphics[width=0.7\textwidth]{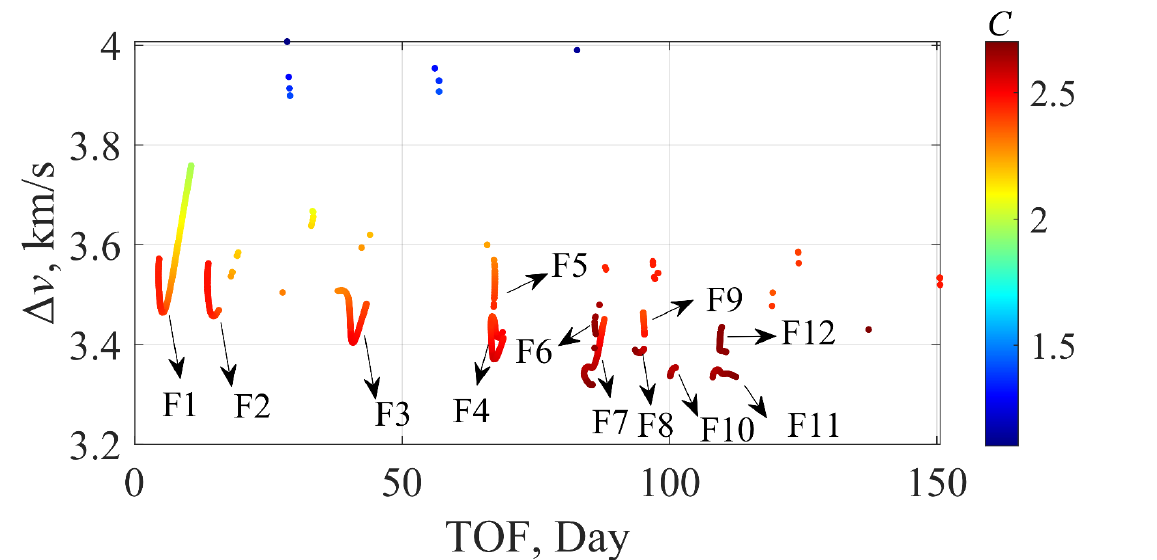}
\caption{The $\left(\text{TOF},\text{ }\Delta v\right)$ map of the obtained solutions.}
\label{fig_global}
\end{figure}

Through correction and continuation mentioned in Section \ref{sec3}, we obtain $5663373$ solutions in total. The $\left(\text{TOF},\text{ }\Delta v\right)$ map of the obtained solutions are shown in Fig. \ref{fig_global}. From Fig. \ref{fig_global}, it is observed that all the values of the obtained $\Delta v$ are less than $4.1\text{ km/s}$. The blank region in the figure is possibly because we only continue the trajectories with $\Delta v<3.45\text{ km/s}$. Moreover, to obtain more transfer families, a more refined grid search and trajectory continuation can be applied. In Fig. \ref{fig_global}, there are 12 typical transfer families, labeled with F1 to F12. These 12 transfer families are manually extracted and the corresponding data can be found in \textbf{supplementary materials}. Subsequently, the distribution of the construction parameters and characteristics of these 12 families are discussed.
\subsection{Families of LEO-DPO Transfers}\label{subsec4.2}
In this subsection, we perform a detailed discussion on the families of obtained LEO-DPO transfers, including the distributions of the construction parameters and characteristics. Specifically, the ranges of the construction parameters are presented to provide a reference to the construction of LEO-DPO transfers, and the ranges of TOF and impulses are presented to provide a reference to the selecting the suitable transfers satisfying the practical mission requirements. Since $\Delta v_i$ typically depends on the capability of the launch vehicle engines while $\Delta v_f$ primarily depends on the fuel capacity of the spacecraft engines \citep{Mingotti2012,FU20254993}, the  distributions of $\left(\text{TOF},\text{ }\Delta v_i\right)$ and $\left(\text{TOF},\text{ }\Delta v_f\right)$ are also presented to provide insights into the mission design (selecting parameters of launch vehicle engines and spacecraft engines). 

\begin{figure}[H]
\centering
\includegraphics[width=0.9\textwidth]{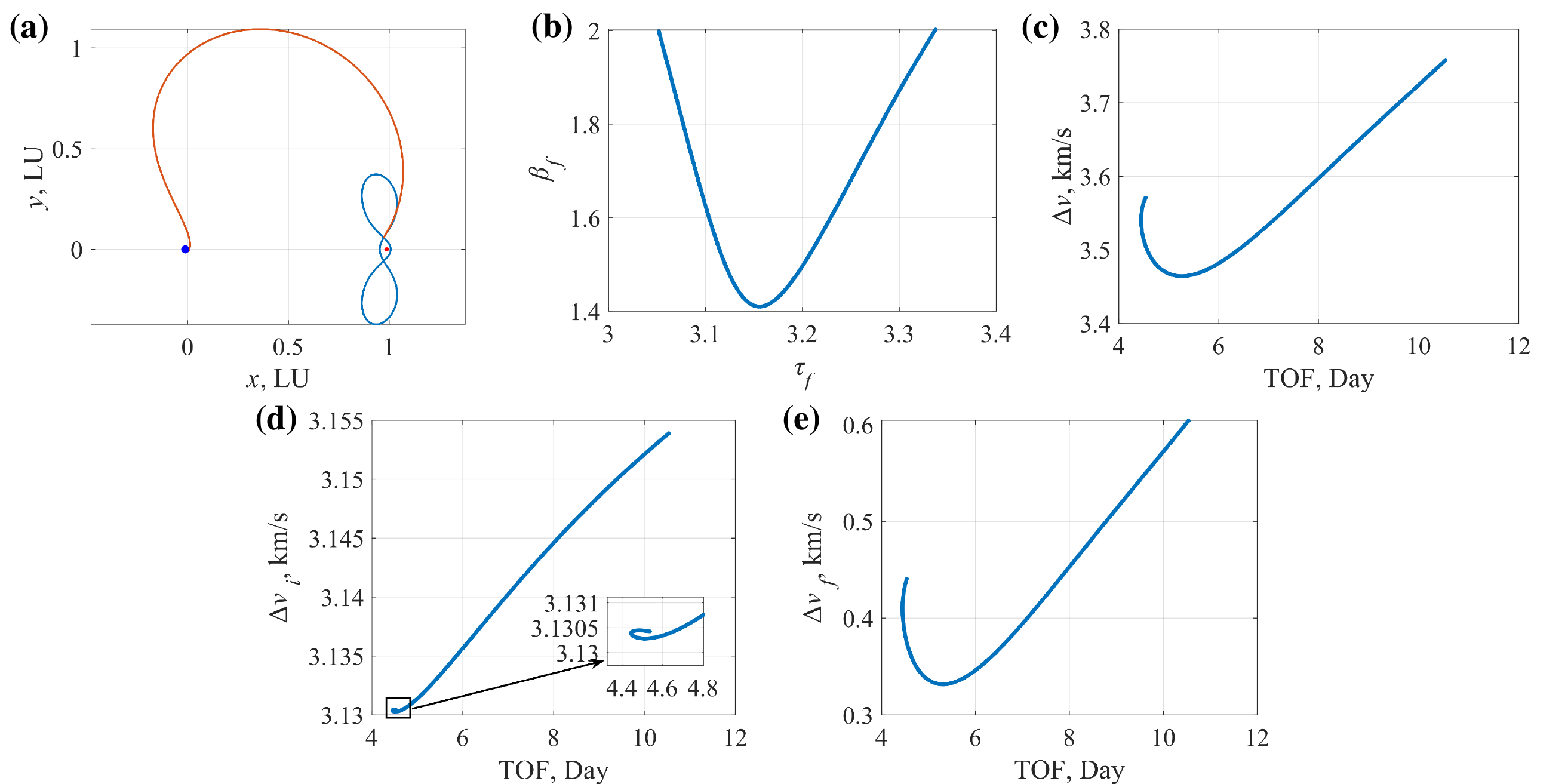}
\caption{Typical trajectory, distributions of $\left(\tau_f,\text{ }\beta_f\right)$, $\left(\text{TOF},\text{ }\Delta v\right)$, $\left(\text{TOF},\text{ }\Delta v_i\right)$, and $\left(\text{TOF},\text{ }\Delta v_f\right)$ for the family F1.}
\label{F1}
\end{figure}

We start with the family F1. The typical trajectory, distributions of $\left(\tau_f,\text{ }\beta_f\right)$, $\left(\text{TOF},\text{ }\Delta v\right)$, $\left(\text{TOF},\text{ }\Delta v_i\right)$, and $\left(\text{TOF},\text{ }\Delta v_f\right)$ are presented in Fig. \ref{F1}. This type of transfers belongs to a typical direct transfer, characterized by relatively short TOF and relatively high $\Delta v$. In the transfer trajectories of this family, the spacecraft does not perform a complete revolution around the Earth. The $\left(\text{TOF},\text{ }\Delta v\right)$ map of this family exhibits a ‘V’ shape, which is similar to the case of Earth-Moon transfers from circular Earth orbits to circular Moon orbits \cite{topputo2013optimal}. Notably, as shown in Fig. \ref{F1} (b), the distribution of $\left(\tau_f,\text{ }\beta_f\right)$ also exhibits a ‘V’ shape. 

\begin{figure}[H]
\centering
\includegraphics[width=0.9\textwidth]{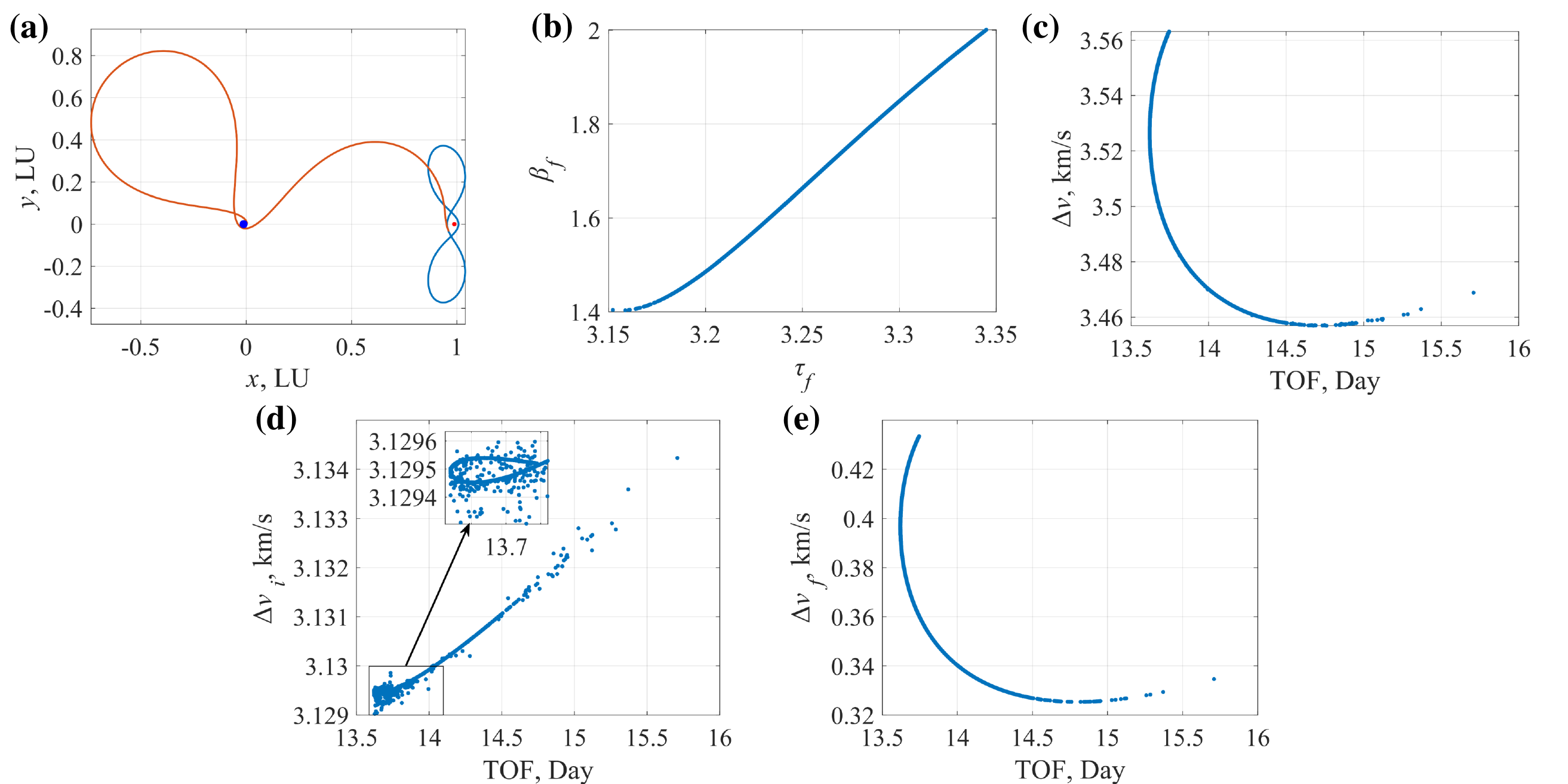}
\caption{Typical trajectory, distributions of $\left(\tau_f,\text{ }\beta_f\right)$, $\left(\text{TOF},\text{ }\Delta v\right)$, $\left(\text{TOF},\text{ }\Delta v_i\right)$, and $\left(\text{TOF},\text{ }\Delta v_f\right)$ for the family F2.}
\label{F2}
\end{figure}

\begin{figure}[H]
\centering
\includegraphics[width=0.9\textwidth]{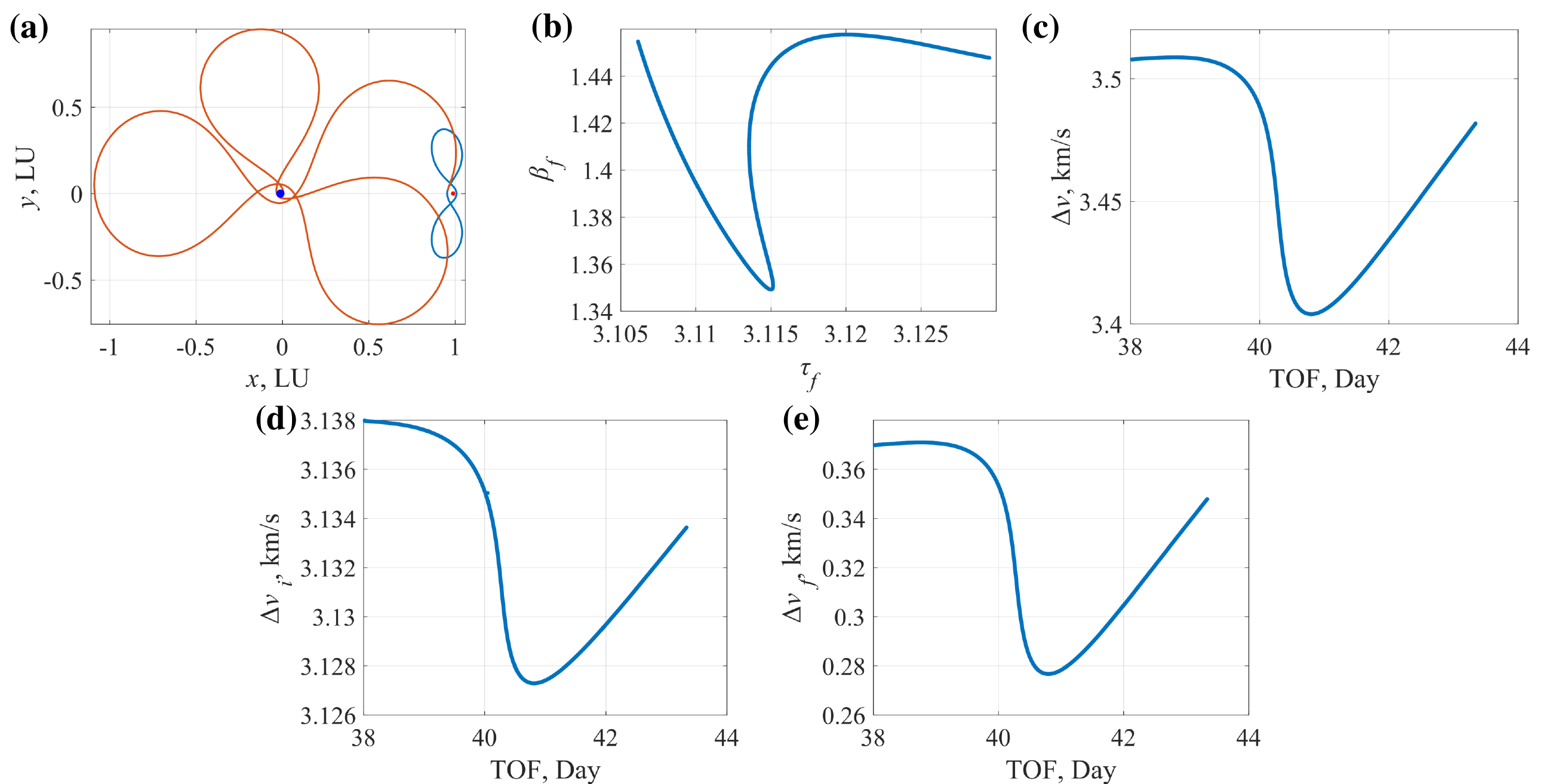}
\caption{Typical trajectory, distributions of $\left(\tau_f,\text{ }\beta_f\right)$, $\left(\text{TOF},\text{ }\Delta v\right)$, $\left(\text{TOF},\text{ }\Delta v_i\right)$, and $\left(\text{TOF},\text{ }\Delta v_f\right)$ for the family F3.}
\label{F3}
\end{figure}

\begin{figure}[H]
\centering
\includegraphics[width=0.9\textwidth]{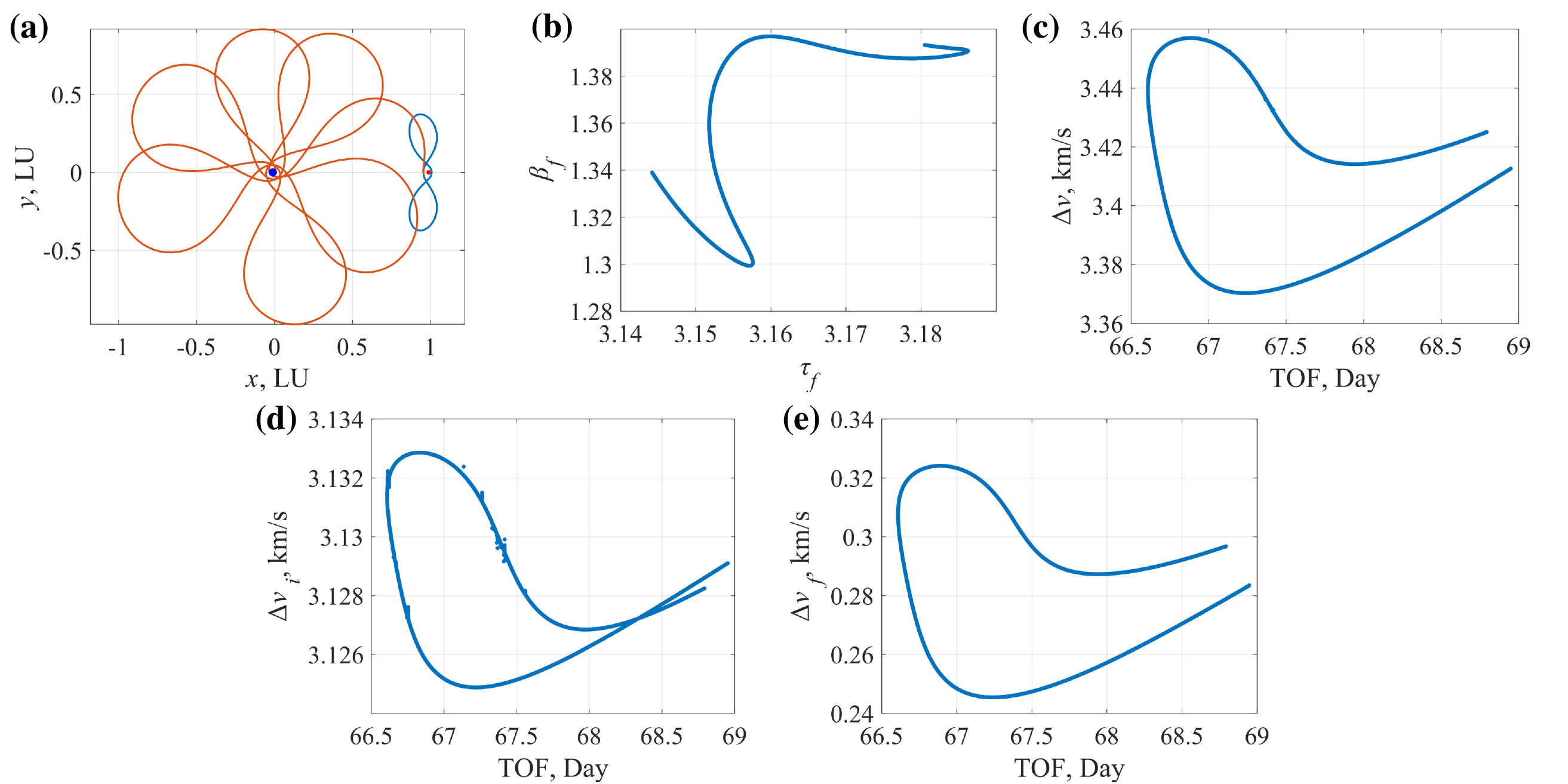}
\caption{Typical trajectory, distributions of $\left(\tau_f,\text{ }\beta_f\right)$, $\left(\text{TOF},\text{ }\Delta v\right)$, $\left(\text{TOF},\text{ }\Delta v_i\right)$, and $\left(\text{TOF},\text{ }\Delta v_f\right)$ for the family F4.}
\label{F4}
\end{figure}

\begin{figure}[H]
\centering
\includegraphics[width=0.9\textwidth]{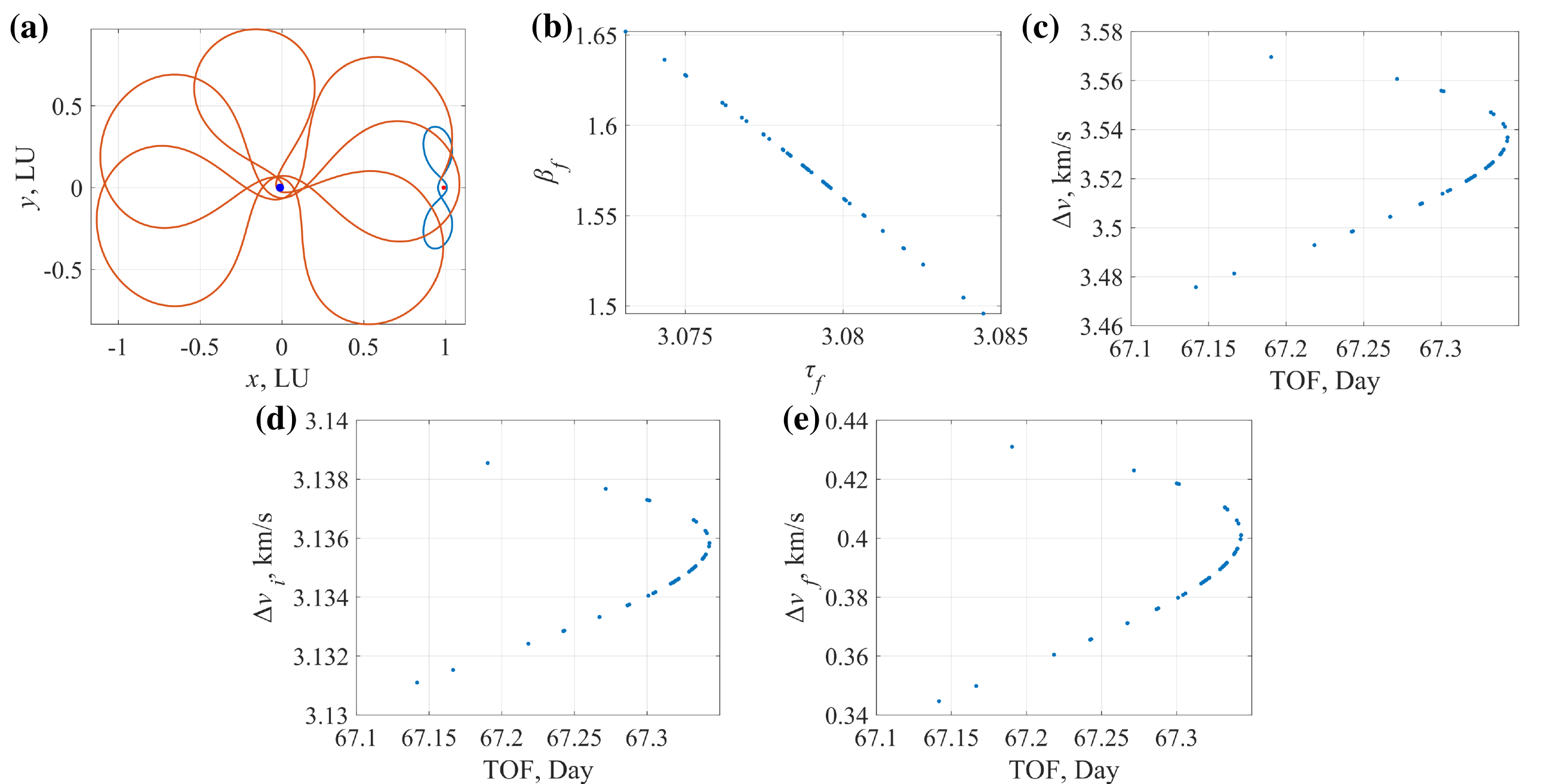}
\caption{Typical trajectory, distributions of $\left(\tau_f,\text{ }\beta_f\right)$, $\left(\text{TOF},\text{ }\Delta v\right)$, $\left(\text{TOF},\text{ }\Delta v_i\right)$, and $\left(\text{TOF},\text{ }\Delta v_f\right)$ for the family F5.}
\label{F5}
\end{figure}

\begin{figure}[H]
\centering
\includegraphics[width=0.9\textwidth]{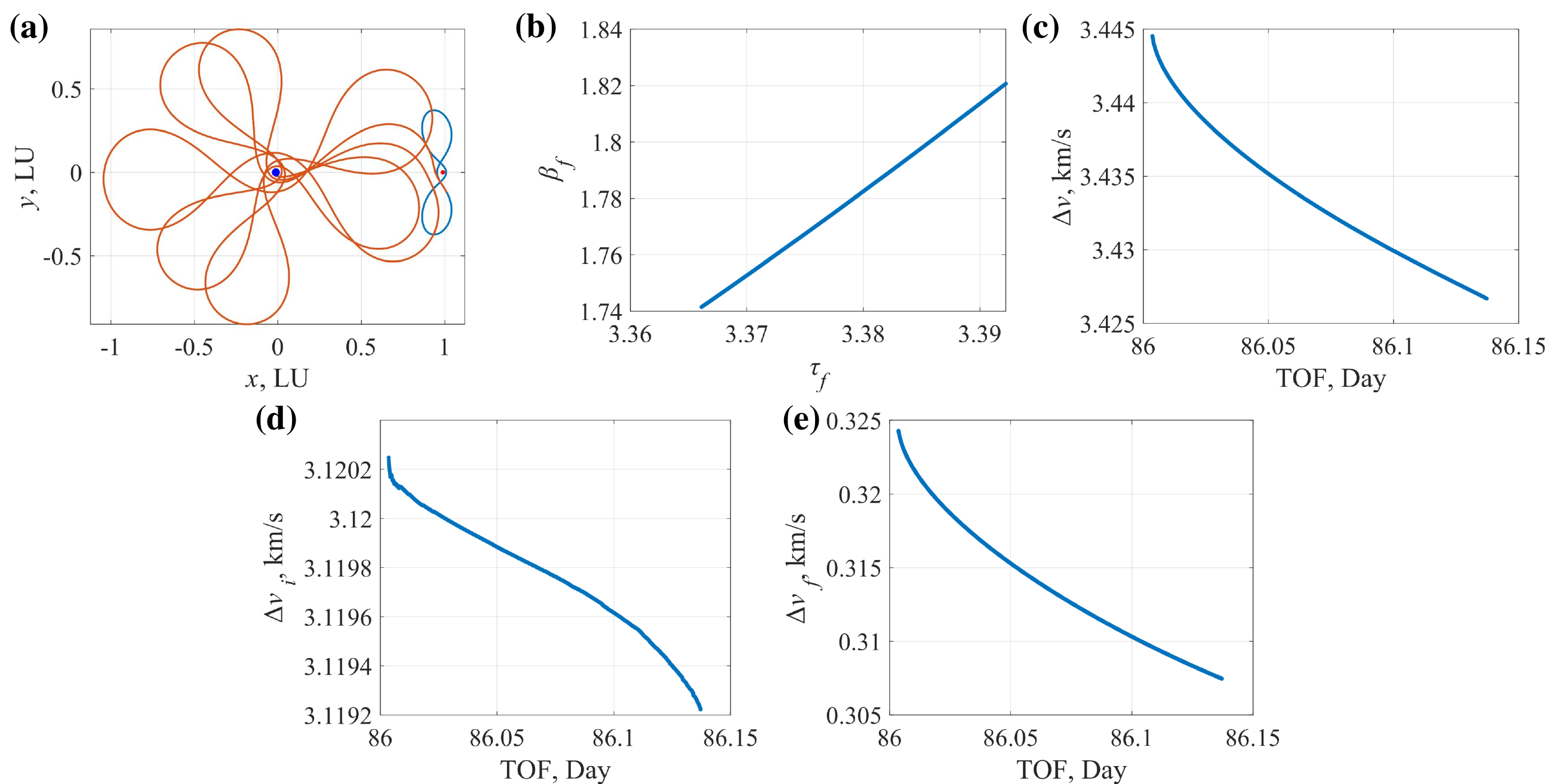}
\caption{Typical trajectory, distributions of $\left(\tau_f,\text{ }\beta_f\right)$, $\left(\text{TOF},\text{ }\Delta v\right)$, $\left(\text{TOF},\text{ }\Delta v_i\right)$, and $\left(\text{TOF},\text{ }\Delta v_f\right)$ for the family F6.}
\label{F6}
\end{figure}

\begin{figure}[H]
\centering
\includegraphics[width=0.9\textwidth]{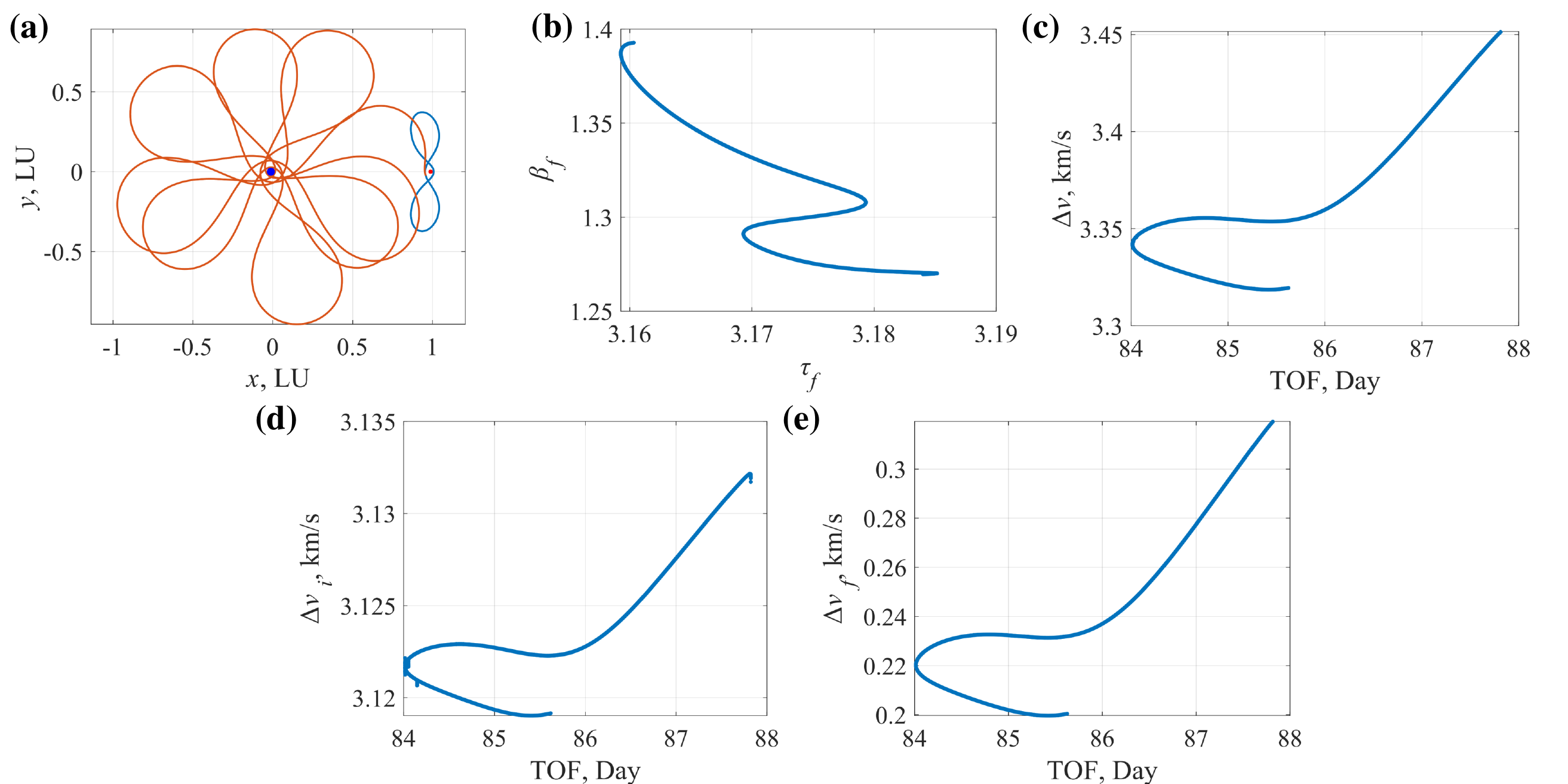}
\caption{Typical trajectory, distributions of $\left(\tau_f,\text{ }\beta_f\right)$, $\left(\text{TOF},\text{ }\Delta v\right)$, $\left(\text{TOF},\text{ }\Delta v_i\right)$, and $\left(\text{TOF},\text{ }\Delta v_f\right)$ for the family F7.}
\label{F7}
\end{figure}

\begin{figure}[H]
\centering
\includegraphics[width=0.9\textwidth]{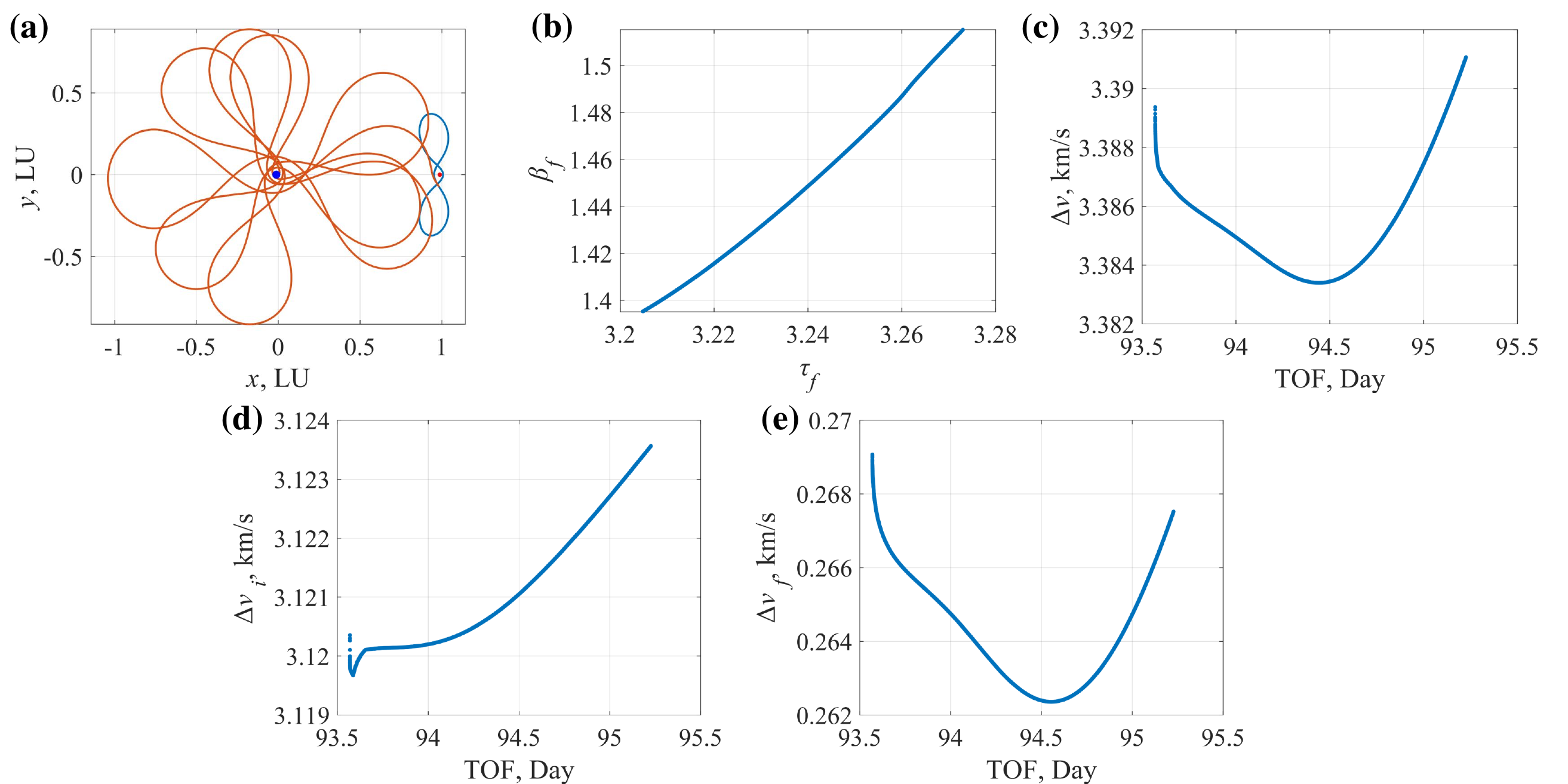}
\caption{Typical trajectory, distributions of $\left(\tau_f,\text{ }\beta_f\right)$, $\left(\text{TOF},\text{ }\Delta v\right)$, $\left(\text{TOF},\text{ }\Delta v_i\right)$, and $\left(\text{TOF},\text{ }\Delta v_f\right)$ for the family F8.}
\label{F8}
\end{figure}

\begin{figure}[H]
\centering
\includegraphics[width=0.9\textwidth]{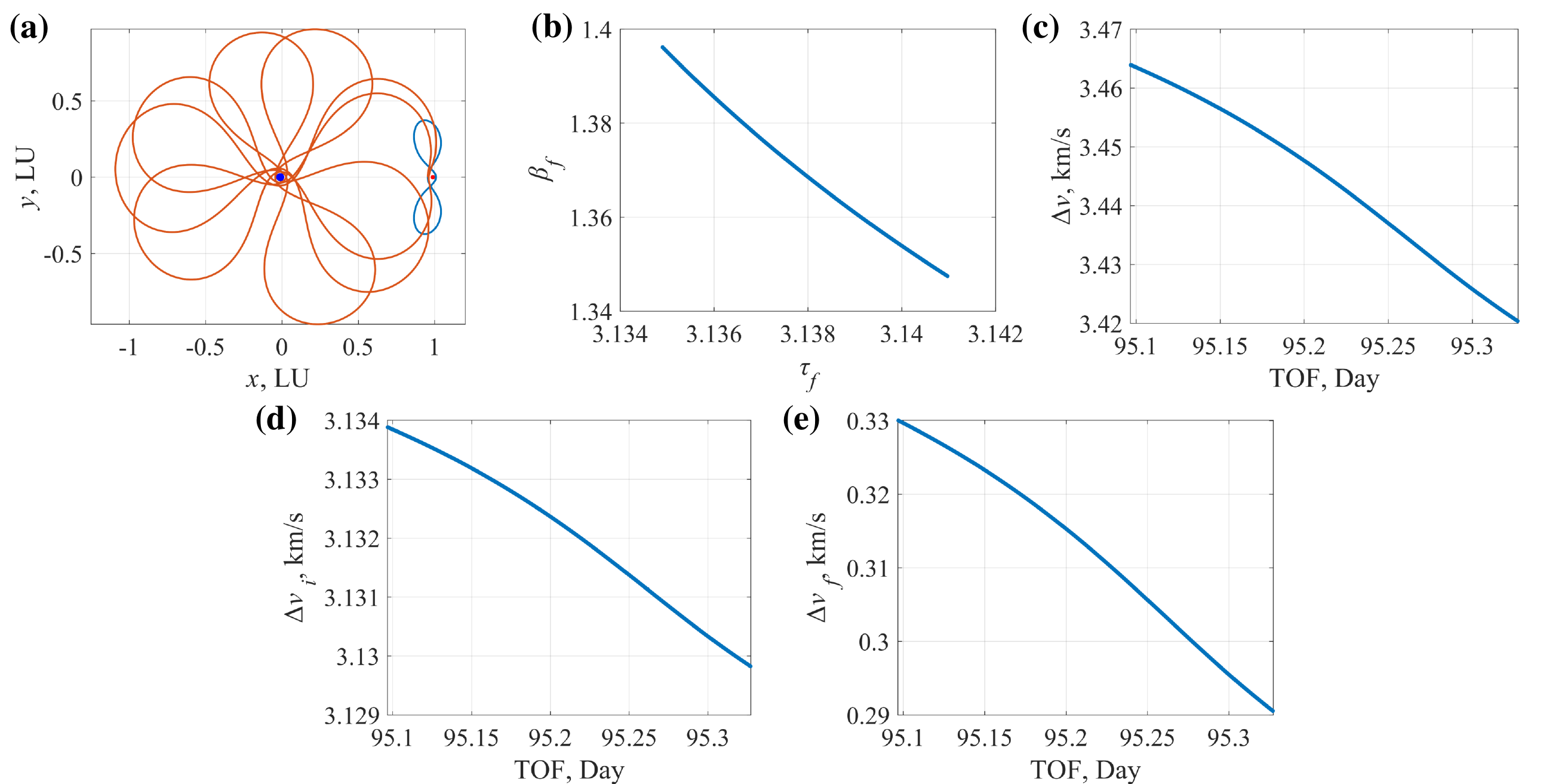}
\caption{Typical trajectory, distributions of $\left(\tau_f,\text{ }\beta_f\right)$, $\left(\text{TOF},\text{ }\Delta v\right)$, $\left(\text{TOF},\text{ }\Delta v_i\right)$, and $\left(\text{TOF},\text{ }\Delta v_f\right)$ for the family F9.}
\label{F9}
\end{figure}

\begin{figure}[H]
\centering
\includegraphics[width=0.9\textwidth]{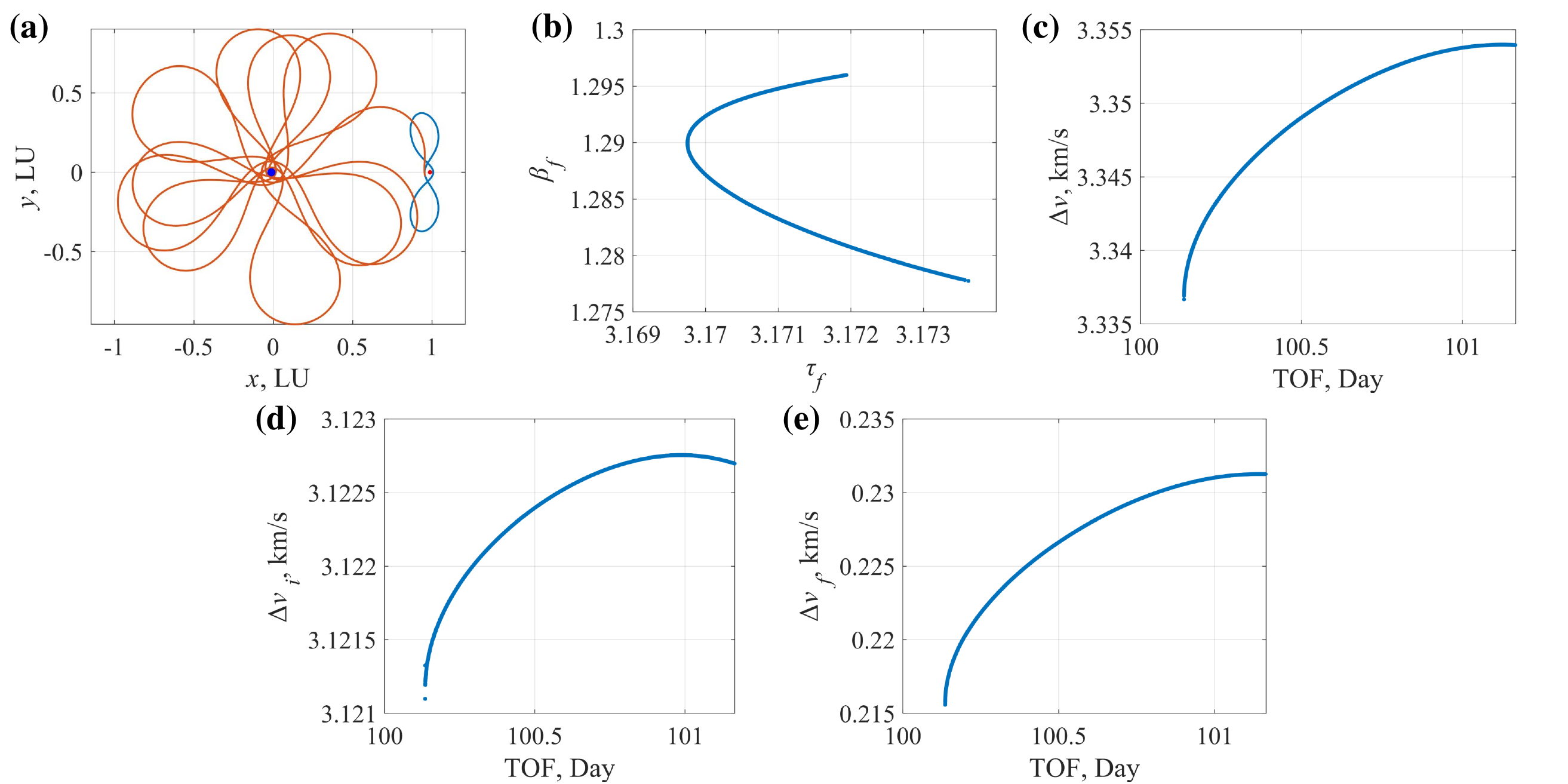}
\caption{Typical trajectory, distributions of $\left(\tau_f,\text{ }\beta_f\right)$, $\left(\text{TOF},\text{ }\Delta v\right)$, $\left(\text{TOF},\text{ }\Delta v_i\right)$, and $\left(\text{TOF},\text{ }\Delta v_f\right)$ for the family F10.}
\label{F10}
\end{figure}

\begin{figure}[H]
\centering
\includegraphics[width=0.9\textwidth]{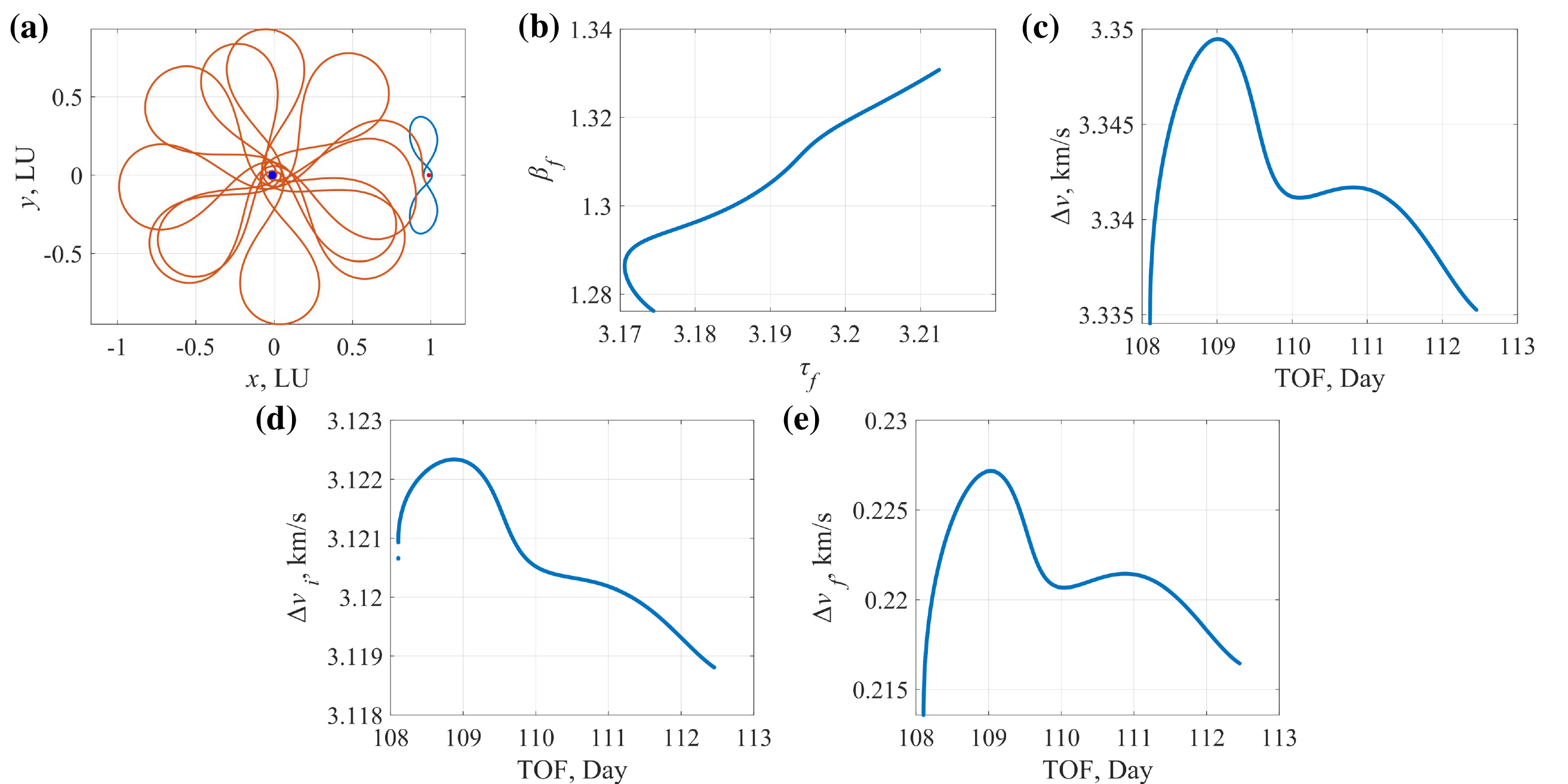}
\caption{Typical trajectory, distributions of $\left(\tau_f,\text{ }\beta_f\right)$, $\left(\text{TOF},\text{ }\Delta v\right)$, $\left(\text{TOF},\text{ }\Delta v_i\right)$, and $\left(\text{TOF},\text{ }\Delta v_f\right)$ for the family F11.}
\label{F11}
\end{figure}

\begin{figure}[H]
\centering
\includegraphics[width=0.9\textwidth]{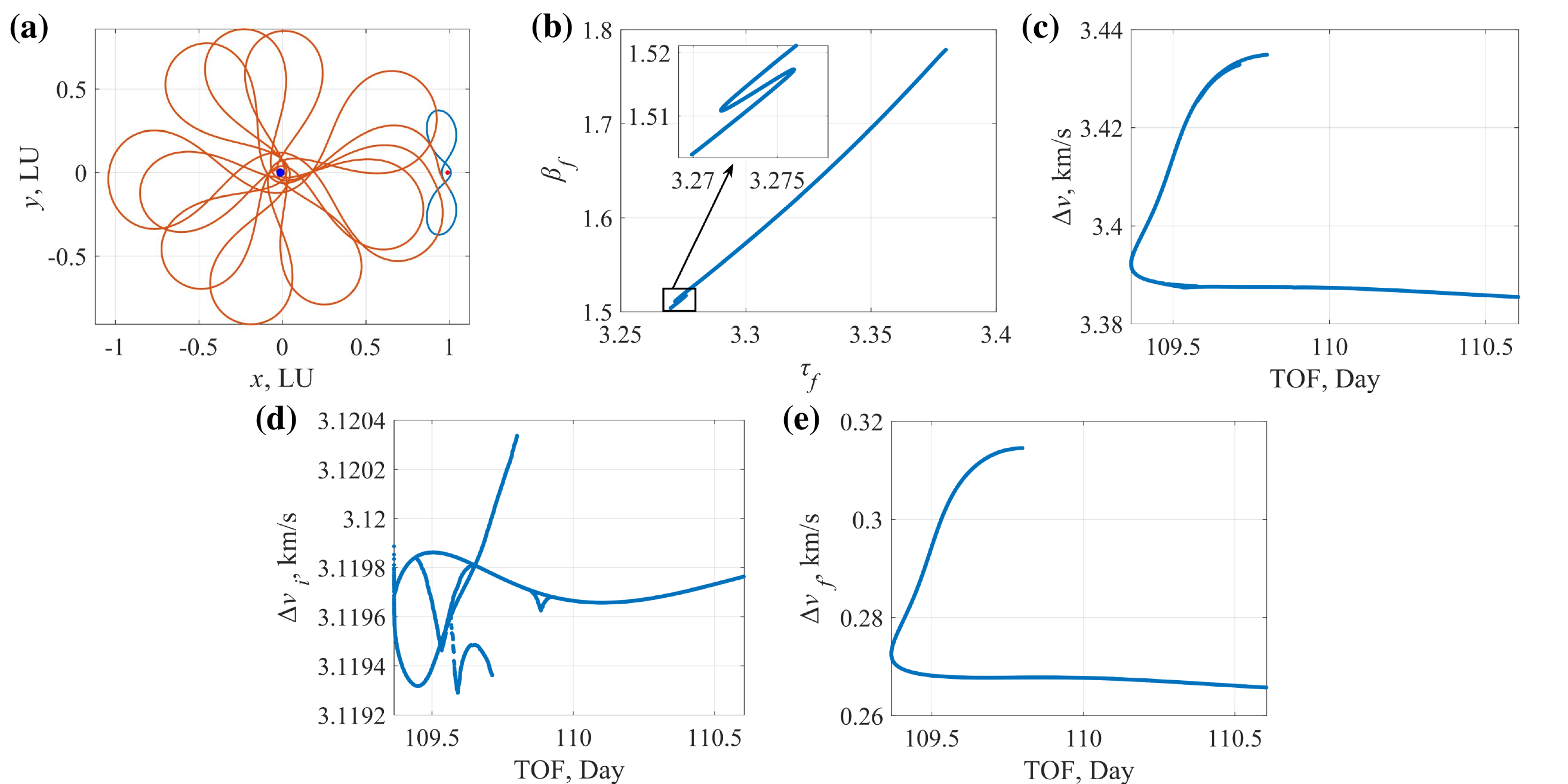}
\caption{Typical trajectory, distributions of $\left(\tau_f,\text{ }\beta_f\right)$, $\left(\text{TOF},\text{ }\Delta v\right)$, $\left(\text{TOF},\text{ }\Delta v_i\right)$, and $\left(\text{TOF},\text{ }\Delta v_f\right)$ for the family F12.}
\label{F12}
\end{figure}

When investigating the family F2, the corresponding trajectory and distributions are presented in Fig. \ref{F2}. In this family, the trajectory preform a complete revolution around the Earth before inserting into the DPO. This type of trajectories are similar to types ‘Family b’ of Earth-Moon transfers from circular Earth orbits to circular Moon orbits in the work of Topputo \cite{topputo2013optimal} in geometry (similar geometry of Earth-Moon transfers from circular Earth orbits to circular Moon orbits can also be found in Refs. \cite{Yagasaki2004a,Yagasaki2004b,mengali2005optimization}). For the distribution of $\left(\text{TOF},\text{ }\Delta v\right)$, $\left(\text{TOF},\text{ }\Delta v_i\right)$, and $\left(\text{TOF},\text{ }\Delta v_f\right)$, it is observed that there exists a scatter distribution in the $\left(\text{TOF},\text{ }\Delta v_i\right)$ map. It is possibly because the constraints $\bm{\psi}_i$ are not satisfied rigorously (i.e., $\left|\left|\bm{\psi}_i\right|\right| \ne \bm{0}$). The scatter distribution in the $\left(\text{TOF},\text{ }\Delta v_i\right)$ map affects the $\left(\text{TOF},\text{ }\Delta v\right)$ map slightly because the variation in $\Delta v_f$ is more significant than that in $\Delta v_i$. This family is also characterized by relatively short TOF and relatively high $\Delta v$. Specifically, the TOF of this family are longer than that of the family F1, as shown in Fig. \ref{fig_global}. 
For the family F3, the corresponding information is presented in Fig. \ref{F3}. It is observed that the trajectory of this family performs three revolutions around the Earth before perform an insertion into the DPO. Compared to the families F1 and F2, an overall reduction of $\Delta v$ and $\Delta v_f$ is observed. However, the TOF of this family exhibits longer than that of the families F1 and F2.
For the family F4, the corresponding information is presented in Fig. \ref{F4}. In this family, the trajectory perform six revolutions around the Earth. A further reduction of $\Delta v$ and $\Delta v_f$ compared to the families F1-F3 is observed, while the TOF exhibits longer than the families F1-F3.
For the family F5, the corresponding information is presented in Fig. \ref{F5}. In this family, the trajectory performs five revolutions around the Earth. As shown in Fig. \ref{fig_global}, $\Delta v$ of this family is higher than that of the family F4. The TOF of this family is comparable to that of the family F4, while longer than the families F1-F3.
For the families F6-F12, these families are characterized by relatively long TOF and relatively low $\Delta v$. The information of these families are shown in Figs. \ref{F6}-\ref{F12}. Among these families, the numbers of revolutions around the Earth performed by the trajectories of these families are: eight for the family F6, eight for the family F7, nine for the family F8, eight for the family F9, ten for the family F10, eleven for the family F11, and eleven for the family F12. Families with the same revolution number possibly belong to the same family. However, with the grid search and trajectory continuation performed in this paper, the bridge between these families has not been found and its search will be the focus of our future work. The distributions of $\left(\tau_f,\text{ }\beta_f\right)$, $\left(\text{TOF},\text{ }\Delta v\right)$, $\left(\text{TOF},\text{ }\Delta v_i\right)$, and $\left(\text{TOF},\text{ }\Delta v_f\right)$ for these families are all presented in figures. It is observed that the values of $\Delta v_i$ of these families are comparable to those of the families F1-F5, while the reduction of $\Delta v$ is mainly caused by the reduction of $\Delta v_f$. After a description of the obtained transfer families, we perform a summary and discussion on the characteristics of transfer families and the link with the previous works.

\subsection{Summary And Discussion}\label{subsec4.3}
\begin{figure}[H]
\centering
\includegraphics[width=0.3\textwidth]{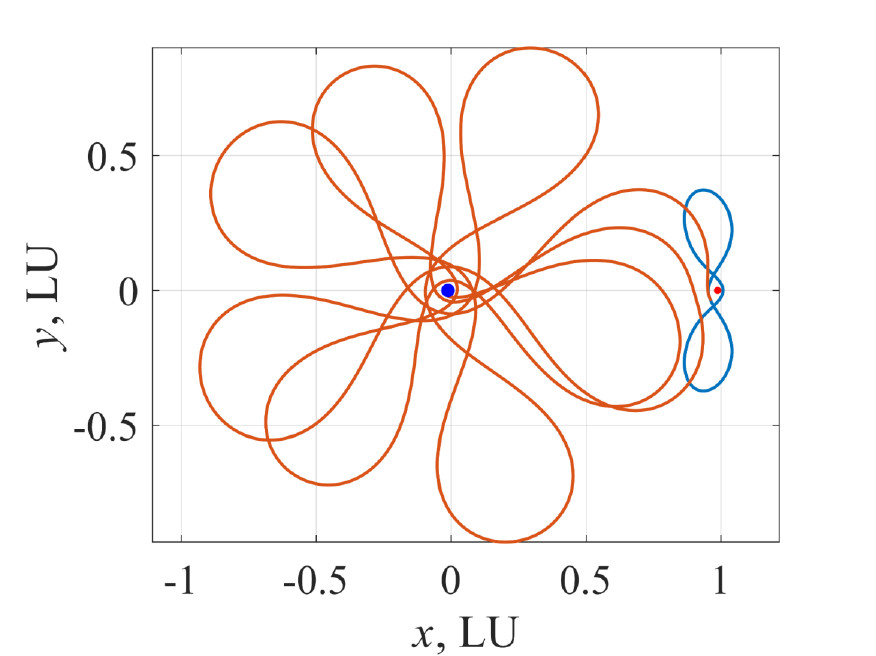}
\caption{Trajectory with the minimum $\Delta v$.}
\label{trajectory}
\end{figure}

Based on the aforementioned discussion, we find that all of our obtained transfer families belong to interior transfer following Topputo's definition \cite{topputo2013optimal}. These results indicate that when using the Earth-Moon PCR3BP to construct the LEO-DPO transfers, most of the obtained transfers belong to interior transfer, not exterior transfer whose apogee exceeds 3-4 times of Earth-Moon distance. We summarize the ranges of $\Delta v$, $\Delta v_i$, $\Delta v_f$, and TOF of the obtained families in Table \ref{tab_summary}, facilitating selection of the transfers suitable for practical missions (to perform a comparison between our obtained solutions and solutions obtained by Mingotti et al. \cite{mingotti2012transfers}, we adopt the same significant figures adopted by them). Among these transfer families, if the practical missions require the fast transfers (e.g., crewed missions), the trajectories of the families F1 and F2 can be selected. When considering a source transport missions which requires a relatively low fuel consumption, the trajectories of the families F4, F6-F12 are suitable. We focus on the solutions with relatively low $\Delta v$. We find that the solution with the minimum $\Delta v$ is located in the family F7, and the corresponding transfer trajectory (denoted as Sample I) is presented in Fig. \ref{trajectory}. In Fig. \ref{trajectory}, it is observed that the trajectory performs a high-altitude lunar flyby \cite{oshima2019low} in the transfer, which helps reduce the $\Delta v$. However, in this paper, we do not find a solution with comparable $\Delta{v}_f$ to the solution with $\Delta{v}_f=0$ obtained by Mingotti et al. \cite{mingotti2012transfers}, as shown in Table \ref{tab_comparison} (Solution I and Solution II). Their solutions were obtained based on the Sun-Earth/Moon planar bicircular restricted four-body problem (PBCR4BP). They firstly computed the invariant manifold of the 1:1 DPO, and patched it with the invariant manifold in the Sun-Earth PCR3BP to generate the initial guesses. The initial guesses were then optimized into the Sun-Earth/Moon PBCR4BP and resulted in exterior transfers with single impulse (i.e., $\Delta v_f=0$). When compared to these solutions, it is observed that $\Delta v_i$ of our solution is lower than that of their solutions. The main difference in $\Delta v$ between our solution and their solutions is due to $\Delta v_f$. Since the solar gravity perturbation is absent in this paper, the single-impulse transfer is not achieved because only dominated by the gravities from the Earth and Moon, the invariant manifolds of the DPO cannot reach the vicinity of the Earth and intersect the considered LEO \cite{mingotti2012transfers}. Moreover, the TOF of our solution is slightly shorter than that of the solutions developed by Mingotti et al. \cite{mingotti2012transfers}. The further exploration of the transfer families in the Earth/Moon PCR3BP and the transfer families constructed in the Sun-Earth/Moon PBCR4BP will be the focus of our future work. The effects of the solar gravity perturbation will be analyzed, and the use of artificial intelligence \cite{pinelli2023neural,zheng2025model,qu2025experience} in the trajectory construction will be explored. 

\begin{table}[!htb]
\caption{Ranges of $\Delta v$, $\Delta v_i$, $\Delta v_f$, and TOF of the obtained families}\label{tab_summary}%
\centering
\renewcommand{\arraystretch}{1.5}
\begin{tabular}{@{}lllll@{}}
\hline
Family  &  $\Delta v$, km/s &  $\Delta v_i$, km/s&  $\Delta v_f$, km/s & TOF, Day \\
\hline
F1     & $3.464-3.758$ & $3.130-3.154$ & $0.332-0.605$   & $4-11$   \\
F2     & $3.457-3.563$ & $3.129-3.134$ & $0.325-0.434$   & $14-16$   \\
F3      & $3.404-3.509$ & $3.127-3.138$ & $0.277-0.371$ & $38-43$    \\
F4      & $3.370-3.457$ & $3.125-3.133$ & $0.245-0.323$ & $67-69$    \\
F5      & $3.476-3.570$ & $3.131-3.139$ & $0.345-0.431$ & $67-67$    \\
F6      & $3.427-3.445$ & $3.119-3.120$ & $0.308-0.324$ & $86-86$    \\
F7      & $3.319-3.452$ & $3.119-3.132$ & $0.200-0.319$ & $84-87$    \\
F8      & $3.383-3.391$ & $3.120-3.124$ & $0.262-0.269$ & $94-95$    \\
F9      & $3.420-3.464$ & $3.130-3.134$ & $0.291-0.330$ & $95-95$    \\
F10      & $3.337-3.354$ & $3.121-3.123$ & $0.216-0.231$ & $100-101$    \\
F11      & $3.335-3.350$ & $3.119-3.122$ & $0.214-0.227$ & $108-112$    \\
F12      & $3.386-3.435$ & $3.119-3.120$ & $0.266-0.315$ & $109-111$    \\
\hline
\end{tabular}
\end{table}

\begin{table}[!htb]
\caption{Link with solutions obtained from the previous works}\label{tab_comparison}%
\centering
\renewcommand{\arraystretch}{1.5}
\begin{tabular}{@{}lllll@{}}
\hline
Solution  &  $\Delta v$, km/s &  $\Delta v_i$, km/s&  $\Delta v_f$, km/s & TOF, Day \\
\hline
Sample I     & $3.319$ & $3.119$ & $0.200$   & $85$   \\
Solution I & $3.161$ & $3.161$ & $0$ & $92$ \\
Solution II & $3.154$ & $3.154$ & $0$ & $90$ \\
\hline
\end{tabular}
\end{table}

\section{Conclusion}\label{sec5}
This paper explores the families of transfers from circular low Earth orbit (LEO) to distant prograde orbit (DPO) in the framework of the Earth-Moon planar circular restricted three-body problem (PCR3BP). Grid search and trajectory continuation are performed to obtain solutions satisfying constraints. Firstly, the states of transfer trajectories are selected in the DPO and a backward strategy is adopted to generate the initial guesses. Then, initial guesses are corrected. Based on the obtained solutions, the linear predictor is derived to predict more feasible solutions in the predictor-corrector continuation method. Through the trajectory continuation, the solution space of transfers is further extended. As a result, we totally obtain 12 families of the considered transfer scenario, most of which are new or less-known compared to the previous works. We perform a discussion on the distributions of construction parameters and characteristics of each family, showing which families are applicable to which specific practical missions. Furthermore, we perform a comparison between our obtained solution with the minimum impulse and solutions obtained from the previous work. Comparison illustrates that our obtained solution advantages slightly in time of flight and Earth injection impulse but fails in total impulse and Moon insertion impulse possibly because of the effect of absence of the solar gravity perturbation. This comparison indicates that there is a potential of using the Sun-Earth/Moon planar restricted four-body problem (PBCR4BP) to construct transfer for further reduction in fuel consumption. Our future work will include the further exploration of the solution space of this transfer scenario in the Earth-Moon PCR3BP, the exploration of the solution space in the Sun-Earth/Moon PBCR4BP, and detailed analysis on the effects of the solar gravity perturbation.

\section*{Declaration of competing interest}
The authors declare that they have no known competing financial interests or personal relationship that could have appeared to influence the work reported in this paper.

\section*{Acknowledgements}

The authors acknowledge the financial support from the National Natural Science Foundation of China (Grant No. 12372044), the National Natural Science Foundation of China (No. U23B6002), the National Natural Science Foundation of China (Grant No. 12302058), the Postdoctoral Science Foundation of China (Grant No. 2024T170480), and the Beijing Natural Science Foundation (No. L252107).

\bibliographystyle{elsarticle-num}

\begin{thebibliography}{10}
\expandafter\ifx\csname url\endcsname\relax
  \def\url#1{\texttt{#1}}\fi
\expandafter\ifx\csname urlprefix\endcsname\relax\def\urlprefix{URL }\fi
\expandafter\ifx\csname href\endcsname\relax
  \def\href#1#2{#2} \def\path#1{#1}\fi

\bibitem{Batcha2020}
A.~L. Batcha, J.~Williams, T.~F. Dawn, J.~P. Gutkowski, M.~V. Widner, S.~L. Smallwood, B.~J. Killeen, E.~C. Williams, R.~E. Harpold, Artemis i trajectory design and optimization, in: 2020 AAS/AIAA Astrodynamics Specialist Conference, no. AAS 20-649, 2020.

\bibitem{Zheng2023}
Y.~Zheng, Y.~Mengfei, D.~Xiangjin, J.~Shengyi, P.~Jing, S.~Yan, G.~Zheng, C.~Liping, P.~Yong, N.~Zhang, Analysis of chang’e-5 lunar core drilling process, Chin. J. Aeronaut. 36~(2) (2023) 292--303.
\newblock \href {https://doi.org/https://doi.org/10.1016/j.cja.2022.01.023} {\path{doi:https://doi.org/10.1016/j.cja.2022.01.023}}.

\bibitem{shetty2025rover}
A.~A. Shetty, H.~Kumar, R.~Regulavalasa, A.~Kamboj, K.~Prakasha, R.~P. HM, G.~Sharma, A.~K. Gupta, A.~Kumar, et~al., Rover ramp deployment system for chandrayaan-3: Concept, design, development and operations, Acta Astronaut. 229 (2025) 1--12.
\newblock \href {https://doi.org/https://doi.org/10.1016/j.actaastro.2024.12.014} {\path{doi:https://doi.org/10.1016/j.actaastro.2024.12.014}}.

\bibitem{Song2023}
Y.-J. Song, J.~Bang, J.~Bae, S.~Hong, Lunar orbit acquisition of the korea pathfinder lunar orbiter: design reference vs actual flight results, Acta Astronaut. 213 (2023) 336--343.
\newblock \href {https://doi.org/https://doi.org/10.1016/j.actaastro.2023.09.021} {\path{doi:https://doi.org/10.1016/j.actaastro.2023.09.021}}.

\bibitem{qi2016minimum}
Y.~Qi, S.~Xu, Minimum $\delta$v for the transfer to permanent lunar orbits with hyperbolic approach, Acta Astronaut. 119 (2016) 183--195.
\newblock \href {https://doi.org/https://doi.org/10.1016/j.actaastro.2015.11.016} {\path{doi:https://doi.org/10.1016/j.actaastro.2015.11.016}}.

\bibitem{qi2016earth}
Y.~Qi, S.~Xu, Earth-moon transfer with near-optimal lunar capture in the restricted four-body problem, Aerosp. Sci. Technol. 55 (2016) 282--291.
\newblock \href {https://doi.org/https://doi.org/10.1016/j.ast.2016.06.008} {\path{doi:https://doi.org/10.1016/j.ast.2016.06.008}}.

\bibitem{Battin1999}
R.~H. Battin, An introduction to the mathematics and methods of astrodynamics, AIAA, 1999.
\newblock \href {https://doi.org/https://doi.org/10.2514/4.861543} {\path{doi:https://doi.org/10.2514/4.861543}}.

\bibitem{Yagasaki2004a}
K.~Yagasaki, Computation of low energy earth-to-moon transfers with moderate flight time, Physica D 197~(3-4) (2004) 313--331.
\newblock \href {https://doi.org/https://doi.org/10.1016/j.physd.2004.07.005} {\path{doi:https://doi.org/10.1016/j.physd.2004.07.005}}.

\bibitem{Yagasaki2004b}
K.~Yagasaki, Sun-perturbed earth-to-moon transfers with low energy and moderate flight time, Celest. Mech. Dyn. Astron. 90 (2004) 197--212.
\newblock \href {https://doi.org/https://doi.org/10.1007/s10569-004-0406-8} {\path{doi:https://doi.org/10.1007/s10569-004-0406-8}}.

\bibitem{topputo2013optimal}
F.~Topputo, On optimal two-impulse earth-moon transfers in a four-body model, Celest. Mech. Dyn. Astron. 117 (2013) 279--313.
\newblock \href {https://doi.org/https://doi.org/10.1007/s10569-013-9513-8} {\path{doi:https://doi.org/10.1007/s10569-013-9513-8}}.

\bibitem{oshima2017analysis}
K.~Oshima, F.~Topputo, S.~Campagnola, T.~Yanao, Analysis of medium-energy transfers to the moon, Celest. Mech. Dyn. Astron. 127 (2017) 285--300.
\newblock \href {https://doi.org/https://doi.org/10.1007/s10569-016-9727-7} {\path{doi:https://doi.org/10.1007/s10569-016-9727-7}}.

\bibitem{oshima2019low}
K.~Oshima, F.~Topputo, T.~Yanao, Low-energy transfers to the moon with long transfer time, Celest. Mech. Dyn. Astron. 131 (2019) 1--19.
\newblock \href {https://doi.org/https://doi.org/10.1007/s10569-019-9883-7} {\path{doi:https://doi.org/10.1007/s10569-019-9883-7}}.

\bibitem{belbruno1993sun}
E.~A. Belbruno, J.~K. Miller, Sun-perturbed earth-to-moon transfers with ballistic capture, J. Guid. Control Dyn. 16~(4) (1993) 770--775.
\newblock \href {https://doi.org/https://doi.org/10.2514/3.21079} {\path{doi:https://doi.org/10.2514/3.21079}}.

\bibitem{campana2025ephemeris}
C.~T. Campana, F.~Topputo, Ephemeris refinement of low-energy earth--moon transfers via an astrodynamic model chain, Acta Astronaut. 232 (2025) 271--282.
\newblock \href {https://doi.org/https://doi.org/10.1016/j.actaastro.2025.03.010} {\path{doi:https://doi.org/10.1016/j.actaastro.2025.03.010}}.

\bibitem{short2014lagrangian}
C.~R. Short, K.~C. Howell, Lagrangian coherent structures in various map representations for application to multi-body gravitational regimes, Acta Astronaut. 94~(2) (2014) 592--607.
\newblock \href {https://doi.org/https://doi.org/10.1016/j.actaastro.2013.08.020} {\path{doi:https://doi.org/10.1016/j.actaastro.2013.08.020}}.

\bibitem{Parker2014}
J.~S. Parker, R.~L. Anderson, Low-energy lunar trajectory design, Vol.~12, John Wiley \& Sons, 2014.

\bibitem{oshima2025graph}
K.~Oshima, A graph-based framework of low-energy transfer design, Acta Astronaut. 229 (2025) 644--661.
\newblock \href {https://doi.org/https://doi.org/10.1016/j.actaastro.2025.01.050} {\path{doi:https://doi.org/10.1016/j.actaastro.2025.01.050}}.

\bibitem{zhang2015transfer}
Z.~Zhang, X.~Hou, Transfer orbits to the earth-moon triangular libration points, Adv. Space Res. 55 (2015) 2899--2913.
\newblock \href {https://doi.org/https://doi.org/10.1016/j.asr.2015.03.008} {\path{doi:https://doi.org/10.1016/j.asr.2015.03.008}}.

\bibitem{tan2020single}
M.~Tan, K.~Zhang, J.~Wang, Single impulsive transfer to the earth--moon triangular point l4 in a bicircular model, Commun. Nonlinear Sci. Numer. Simul. 82 (2020) 105074.
\newblock \href {https://doi.org/https://doi.org/10.1016/j.cnsns.2019.105074} {\path{doi:https://doi.org/10.1016/j.cnsns.2019.105074}}.

\bibitem{demeyer2007transfer}
J.~Demeyer, P.~Gurfil, Transfer to distant retrograde orbits using manifold theory, J. Guid. Control Dyn. 30~(5) (2007) 1261--1267.
\newblock \href {https://doi.org/https://doi.org/10.2514/1.24960} {\path{doi:https://doi.org/10.2514/1.24960}}.

\bibitem{capdevila2018transfer}
L.~R. Capdevila, K.~C. Howell, A transfer network linking earth, moon, and the triangular libration point regions in the earth-moon system, Adv. Space Res. 62~(7) (2018) 1826--1852.
\newblock \href {https://doi.org/https://doi.org/10.1016/j.asr.2018.06.045} {\path{doi:https://doi.org/10.1016/j.asr.2018.06.045}}.

\bibitem{zhang2021transfers}
R.~Zhang, Y.~Wang, C.~Zhang, H.~Zhang, The transfers from lunar dros to earth orbits via optimization in the four body problem, Astrophys. Space Sci. 366~(6) (2021) 49.
\newblock \href {https://doi.org/https://doi.org/10.1007/s10509-021-03955-1} {\path{doi:https://doi.org/10.1007/s10509-021-03955-1}}.

\bibitem{wang2025mechanism}
W.~Ming, C.~Zhang, H.~Zhang, Mechanism analysis of the dro low-energy transfer problem: An energy perspective, Astrodynamics 9 (2025) 165--193.
\newblock \href {https://doi.org/https://doi.org/10.1007/s42064-024-0215-7} {\path{doi:https://doi.org/10.1007/s42064-024-0215-7}}.

\bibitem{mingotti2012transfers}
G.~Mingotti, F.~Topputo, F.~Bernelli-Zazzera, Transfers to distant periodic orbits around the moon via their invariant manifolds, Acta Astronaut. 79 (2012) 20--32.
\newblock \href {https://doi.org/https://doi.org/10.1016/j.actaastro.2012.04.022} {\path{doi:https://doi.org/10.1016/j.actaastro.2012.04.022}}.

\bibitem{henon1969numerical}
M.~H{\'e}non, Numerical exploration of the restricted problem, v, Astron. Astrophys. 1 (1969) 223--238.

\bibitem{parker2010chaining}
J.~S. Parker, K.~E. Davis, G.~H. Born, Chaining periodic three-body orbits in the earth--moon system, Acta Astronaut. 67~(5-6) (2010) 623--638.
\newblock \href {https://doi.org/https://doi.org/10.1016/j.actaastro.2010.04.003} {\path{doi:https://doi.org/10.1016/j.actaastro.2010.04.003}}.

\bibitem{gupta2021earth}
M.~Gupta, K.~C. Howell, C.~Frueh, Earth-moon multi-body orbits to facilitate cislunar surveillance activities, in: AIAA/AAS Astrodynamics Specialist Conference, 2021, pp. 17--18.

\bibitem{Szebehely1967}
V.~Szebehely, Theory of orbit: The restricted problem of three Bodies, Academic, 1967.

\bibitem{dutt2018review}
P.~Dutt, A review of low-energy transfers, Astrophys. Space Sci. 363~(12) (2018) 253.
\newblock \href {https://doi.org/https://doi.org/10.1007/s10509-018-3461-4} {\path{doi:https://doi.org/10.1007/s10509-018-3461-4}}.

\bibitem{scheuerle2025energy}
S.~T. Scheuerle~Jr, K.~C. Howell, D.~C. Davis, Energy-informed pathways: A fundamental approach to designing ballistic lunar transfers, Adv. Space Res. 75~(1) (2025) 1096--1117.
\newblock \href {https://doi.org/https://doi.org/10.1016/j.asr.2024.07.035} {\path{doi:https://doi.org/10.1016/j.asr.2024.07.035}}.

\bibitem{Koon2001}
W.~S. Koon, M.~W. Lo, J.~E. Marsden, S.~D. Ross, Low energy transfer to the moon, Celest. Mech. Dyn. Astron. 81~(1-2) (2001) 63--73.
\newblock \href {https://doi.org/https://doi.org/10.1023/A:1013359120468} {\path{doi:https://doi.org/10.1023/A:1013359120468}}.

\bibitem{Onozaki2017}
K.~Onozaki, H.~Yoshimura, S.~D. Ross, Tube dynamics and low energy earth--moon transfers in the 4-body system, Adv. Space Res. 60~(10) (2017) 2117--2132.
\newblock \href {https://doi.org/https://doi.org/10.1016/j.asr.2017.07.046} {\path{doi:https://doi.org/10.1016/j.asr.2017.07.046}}.

\bibitem{FU20254993}
S.~Fu, Y.~Liang, D.~Wu, S.~Gong, Low-energy earth-moon transfers with lunar ballistic capture based on lagrangian coherent structures in a four-body model, Adv. Space Res. 75~(6) (2025) 4993--5013.
\newblock \href {https://doi.org/https://doi.org/10.1016/j.asr.2024.12.072} {\path{doi:https://doi.org/10.1016/j.asr.2024.12.072}}.

\bibitem{fu2025four}
S.~Fu, D.~Wu, P.~Shi, S.~Gong, Four-body transit-orbit classification combining lagrangian coherent structures with data mining, J. Guid. Control Dyn. (2025) 1--20\href {https://doi.org/https://doi.org/10.2514/1.G008681} {\path{doi:https://doi.org/10.2514/1.G008681}}.

\bibitem{mccarthy2021leveraging}
B.~P. McCarthy, K.~C. Howell, Leveraging quasi-periodic orbits for trajectory design in cislunar space, Astrodynamics 5~(2) (2021) 139--165.
\newblock \href {https://doi.org/https://doi.org/10.1007/s42064-020-0094-5} {\path{doi:https://doi.org/10.1007/s42064-020-0094-5}}.

\bibitem{FU2025}
S.~Fu, S.~Gong, Escape criterion for restricted three-body problem, Adv. Space Res. (2025).
\newblock \href {https://doi.org/https://doi.org/10.1016/j.asr.2025.02.017} {\path{doi:https://doi.org/10.1016/j.asr.2025.02.017}}.

\bibitem{oshima2021capture}
K.~Oshima, Capture and escape analyses on planar retrograde periodic orbit around the earth, Adv. Space Res. 68~(9) (2021) 3891--3902.
\newblock \href {https://doi.org/https://doi.org/10.1016/j.asr.2021.07.012} {\path{doi:https://doi.org/10.1016/j.asr.2021.07.012}}.

\bibitem{dei2018survey}
D.~A. Dei~Tos, R.~P. Russell, F.~Topputo, Survey of mars ballistic capture trajectories using periodic orbits as generating mechanisms, J. Guid. Control Dyn. 41~(6) (2018) 1227--1242.
\newblock \href {https://doi.org/https://doi.org/10.2514/1.G003158} {\path{doi:https://doi.org/10.2514/1.G003158}}.

\bibitem{du2022transfer}
C.~Du, O.~L. Starinova, Y.~Liu, Transfer between the planar lyapunov orbits around the earth-moon l2 point using low-thrust engine, Acta Astronaut. 201 (2022) 513--525.
\newblock \href {https://doi.org/https://doi.org/10.1016/j.actaastro.2022.09.056} {\path{doi:https://doi.org/10.1016/j.actaastro.2022.09.056}}.

\bibitem{du2023two}
C.~Du, K.~Wu, O.~L. Starinova, Y.~Liu, Two trajectory configurations for the low-thrust transfer between northern and southern halo orbits in the earth-moon system, Adv. Space Res. 72~(10) (2023) 4093--4105.
\newblock \href {https://doi.org/https://doi.org/10.1016/j.asr.2023.08.007} {\path{doi:https://doi.org/10.1016/j.asr.2023.08.007}}.

\bibitem{jiao2024asteroid}
Y.~Jiao, B.~Cheng, Y.~Huang, E.~Asphaug, B.~Gladman, R.~Malhotra, P.~Michel, Y.~Yu, H.~Baoyin, Asteroid kamo ‘oalewa’s journey from the lunar giordano bruno crater to earth 1: 1 resonance, Nat. Astron. 8~(7) (2024) 819--826.
\newblock \href {https://doi.org/https://doi.org/10.1038/s41550-024-02258-z} {\path{doi:https://doi.org/10.1038/s41550-024-02258-z}}.

\bibitem{fu2025constructing}
S.~Fu, D.~Wu, X.~Liu, P.~Shi, S.~Gong, Constructing four-body ballistic lunar transfers via analytical energy conditions, arXiv preprint arXiv:2504.16804 (2025).
\newblock \href {https://doi.org/https://doi.org/10.48550/arXiv.2504.16804} {\path{doi:https://doi.org/10.48550/arXiv.2504.16804}}.

\bibitem{fu2025analytical}
S.~Fu, D.~Wu, S.~Gong, Analytical nonlinear predictor for trajectory continuation in multibody models, J. Guid. Control Dyn. 48~(6) (2025) 1418--1427.
\newblock \href {https://doi.org/https://doi.org/10.2514/1.G008863} {\path{doi:https://doi.org/10.2514/1.G008863}}.

\bibitem{singh2019mission}
S.~K. Singh, E.~Taheri, R.~Woollands, J.~Junkins, Mission design for close-range lunar mapping by quasi-frozen orbits, in: 70th International Astronautical Congress, Washington DC, USA, 2019, pp. 1--11.

\bibitem{pushparaj2021transfers}
N.~Pushparaj, N.~Baresi, Y.~Kawakatsu, Transfers and orbital maintenance of spatial retrograde orbits for phobos exploration, Acta Astronaut. 189 (2021) 452--464.
\newblock \href {https://doi.org/https://doi.org/10.1016/j.actaastro.2021.09.008} {\path{doi:https://doi.org/10.1016/j.actaastro.2021.09.008}}.

\bibitem{singh2021exploiting}
S.~K. Singh, B.~D. Anderson, E.~Taheri, J.~L. Junkins, Exploiting manifolds of l1 halo orbits for end-to-end earth-moon low-thrust trajectory design, Acta Astronaut. 183 (2021) 255--272.
\newblock \href {https://doi.org/https://doi.org/10.1016/j.actaastro.2021.03.017} {\path{doi:https://doi.org/10.1016/j.actaastro.2021.03.017}}.

\bibitem{pushparaj2024optimization}
N.~Pushparaj, N.~Baresi, Y.~Kawakatsu, Optimization of mmx relative quasi-satellite transfer trajectories using primer vector theory, Acta Astronaut. 225 (2024) 390--401.
\newblock \href {https://doi.org/https://doi.org/10.1016/j.actaastro.2024.09.031} {\path{doi:https://doi.org/10.1016/j.actaastro.2024.09.031}}.

\bibitem{Mingotti2012}
G.~Mingotti, F.~Topputo, F.~Bernelli-Zazzera, Efficient invariant-manifold, low-thrust planar trajectories to the moon, Commun. Nonlinear Sci. Numer. Simul. 17~(2) (2012) 817--831.
\newblock \href {https://doi.org/https://doi.org/10.1016/j.cnsns.2011.06.033} {\path{doi:https://doi.org/10.1016/j.cnsns.2011.06.033}}.

\bibitem{mengali2005optimization}
G.~Mengali, A.~A. Quarta, Optimization of biimpulsive trajectories in the earth-moon restricted three-body system, J. Guid. Control Dyn. 28~(2) (2005) 209--216.
\newblock \href {https://doi.org/https://doi.org/10.2514/1.7702} {\path{doi:https://doi.org/10.2514/1.7702}}.

\bibitem{pinelli2023neural}
A.~Pinelli, Neural network applications to bi-impulsive earth-moon transfers in a four-body model, Master's thesis, Politecnico di Milano (2023).

\bibitem{zheng2025model}
T.~Zheng, L.~Cheng, S.~Gong, X.~Huang, Model incremental learning of flight dynamics enhanced by sample management, Aerosp. Sci. Technol. 160 (2025) 110049.
\newblock \href {https://doi.org/https://doi.org/10.1016/j.ast.2025.110049} {\path{doi:https://doi.org/10.1016/j.ast.2025.110049}}.

\bibitem{qu2025experience}
C.~Qu, L.~Cheng, S.~Gong, X.~Huang, Experience replay enhances excitation condition of neural-network adaptive control learning, J. Guid. Control Dyn. 48~(3) (2025) 496--507.
\newblock \href {https://doi.org/https://doi.org/10.2514/1.G008162} {\path{doi:https://doi.org/10.2514/1.G008162}}.

\end{thebibliography}


\printcredits

\end{document}